# Self-Assembly of Delta-Formamidinium Lead Iodide Nanoparticles to Nanorods: Study of Memristor Properties and Resistive Switching Mechanism

*Chinnadurai Muthu, A. N. Resmi, Avija Ajayakumar, N. E. Aswathi Ravindran, G. Dayal, K. B. Jinesh,* Konrad Szaciłowski, and Chakkooth Vijayakumar**

C. Muthu, A. Ajayakumar, N. E. A. Ravindran, C. Vijayakumar Chemical Sciences and Technology Division
CSIR-National Institute for Interdisciplinary Science and Technology (CSIR-NIIST)
Thiruvananthapuram 695 019, India E-mail: cvijayakumar@niist.res.in
C. Muthu, A. Ajayakumar, C. Vijayakumar
Academy of Scientific and Innovative Research (AcSIR) Ghaziabad 201 002, India
A. N. Resmi, G. Dayal, K. B. Jinesh Department of Physics
Indian Institute of Space Science and Technology (IIST) Thiruvananthapuram 695 547, India
E-mail: kbjinesh@iist.ac.in
K. Szaciłowski
Academic Centre for Materials and Nanotechnology AGH University of Krakow
Mickiewicza 30, Krakow 30 059, Poland

**Abstract**

*In the quest for advanced memristor technologies, this study introduces the synthesis of delta-formamidinium lead iodide (**δ**-FAPbI3) nanoparticles (NPs) and their self-assembly into nanorods (NRs). The formation of these NRs is facilitated by iodide vacancies, promoting the fusion of individual NPs at higher concentrations. Notably, these NRs exhibit robust stability under ambient conditions, a distinctive advantage attributed to the presence of capping ligands and a crystal lattice structured around face-sharing octahedra. When employed as the active layer in resistive random-access memory devices, these NRs demonstrate exceptional bipolar switching properties. A remarkable on/off ratio (105) is achieved, surpassing the performances of previously reported low-dimensional perovskite derivatives and **α**-FAPbI3 NP-based devices. This enhanced performance is attributed to the low off-state current owing to the reduced number of halide vacancies, intrinsic low*

*dimensionality, and the parallel alignment of NRs on the FTO substrate. This study not only provides significant insights into the development of superior materials for memristor applications but also opens new avenues for exploring low-dimensional perovskite derivatives in advanced electronic devices.*

## 1. Introduction

The rapid advancement of information technology has led to a demand for memory devices with high storage density, which is difficult to achieve using traditional silicon-based flash memories as their size has reached its physical limitations.[1] Resistive random access memory (ReRAM) devices have emerged as a promising alternative due to their simple architecture, low power consumption, high operation speed, and long data retention.[2] These devices can store large amounts of data as their size can be reduced to near the ionic diameter level. They also allow nonvolatile data processing and storage in a single device. These advantages have led to increased research on ReRAM devices, with a specific focus on the appropriate selection of active materials to develop high-performance devices. Among these materials, halide perovskites have been found to exhibit excellent resistive switching properties due to the unique current-voltage hysteresis resulting from halide vacancies and their migration under an electric field.[3] While 3D perovskites have shown good device performance, they are known to be sensitive to moisture, heat, and light, which can result in poor stability that limits their practical application.[4]

Low-dimensional perovskite/perovskite-derivative materials exhibit a high surface area and anisotropic conduction compared to bulk 3D perovskites.[5] The hydrophobic nature of organic cations as well as structure based on face-sharing octahedra imparts exceptional environmental stability to the materials. Although large bandgap low-dimensional perovskites are unsuitable for solar cell applications, they possess several advantages for memory devices.[6] For example, the resistance changes in 3D perovskite-based ReRAMs, typically driven by methylammonium ($MA^+$) ion migration, are not present in low-dimensional materials. This absence, attributable to the larger organic cations, makes the devices more reliable.[6c] These materials can efficiently form conductive filaments owing to anisotropic defect migration.[6e] The large bandgap helps reduce the thermal excitation of the electrons, thereby reducing the off current. Additionally, the formation and rupture of conducting filaments formed by halide vacancies in these materials are much easier due to their weak interaction energy, leading to an excellent on/off ratio and improved switching dynamics.[6f,d] Furthermore, the halide ion migration in these materials under an electric field is negligible compared to their corresponding bulk counterparts, leading to reduced off current and improved endurance under continuous electric stimulation.[6b]

Park et al. conducted a detailed study on the effect of dimensionality on the resistive switching (RS) properties of $BA_2(MA)_{n-1}Pb_nI_{3n+1}$ (BA: butylammonium; n = 1, 2, 3, …∞).[6e] The anisotropic $BA_2PbI_4$ exhibited superior RS properties, with the lowest set electric field and a high on/off ratio, compared to quasi-two-dimensional and three-dimensional perovskites. This was attributed to the enhanced Schottky barrier between the perovskite and electrode for the large-bandgap $BA_2PbI_4$. Park's group also analyzed the effect of RS on the interlayer spacing in $[C_6H_5(CH_2)_nNH_3]_2Pb_nI_{3n+1}$.[7] They found that the RS properties were gradually enhanced by increasing the chain length, which was attributed to the improved two-dimensional molecular confinement in $[C_6H_5CH_2NH_3]_2PbI_4$ that facilitates the formation/rupture of filaments. Xia et al. have fabricated two-dimensional/three-dimensional perovskite heterostructures by depositing *n*-butylammonium iodide (BAI) on top of $MAPbI_{3-X}Cl_X$ to enhance the on/off ratio of three-dimensional perovskites.[8] They found that a thin two-dimensional $(BA)_2MA_{n-1}Pb_nI_{3n-1}$ interlayer formed on the surface of the three-dimensional perovskite passivated its defects, thus decreasing the off-current and leading to an improved on/off ratio. However, most of these materials exhibit average RS behavior because of the poor film quality stemming from their polycrystalline nature.

Previously, we demonstrated the memristor characteristics of methylammonium [9] and formamidinium lead halide perovskite nanoparticles (NPs),[10] which performed superior to the corresponding bulk materials. Other research groups have similarly explored nanostructure-based ReRAMs using 3D perovskites and low-dimensional perovskites or perovskite derivatives.[6d,11] However, most of these materials require polymers for stability [6d] or templates for the suitable alignment [11c,d], limiting their practicality. In contrast, the present work reports a polymer-free and template-free synthesis of delta form of $FAPbI_3$ ($\delta$-$FAPbI_3$; a low-dimensional perovskite-derivative) and its self-assembly into nanorods (NRs) with excellent resistive switching characteristics. The $\delta$-$FAPbI_3$ is characterized as a yellow phase material with a hexagonal structure, exhibiting an indirect bandgap and non-luminescent properties.[12] Generally, this material originates from the rapid phase transition of the black, trigonal, and direct bandgap alpha phase ($\alpha$-$FAPbI_3$) under ambient conditions, driven by a reduction in the Pb—I—Pb bond angle, especially in the presence of moisture. Notably, the $\delta$-phase possesses excellent environmental stability because of its face-sharing inorganic $PbI_6^{4-}$ octahedra. Unlike previous studies, the adopted methodology bypasses the synthesis of bulk $\delta$-$FAPbI_3$, a process typically mired with impurities such as $PbI_2$ and $\alpha$-$FAPbI_3$ stemming from the phase transition of $\alpha$-$FAPbI_3$.[13] One of the major challenges with existing perovskite-based ReRAM devices is the high off-current attributed to the abundance of halide ion vacancies and their swift migration under an electric field.[14] These issues result in a decrease in the on/off ratio and degradation of the materials, thereby curtailing the efficiency and longevity of the devices. The novel approach described herein addresses this issue by transitioning $\delta$-$FAPbI_3$ NPs into NRs, significantly curtailing the off-current in ReRAM devices, even with a simple device architecture. This transition is facilitated by an excess of halide ion vacancies, which catalyze the self-assembly of the NPs into NRs, concurrently reducing the concentration of halide vacancies. These NRs, aligned parallel to the FTO substrate without the need for templates, exhibited superior structural integrity and operational efficiency. As a result, $\delta$-$FAPbI_3$ NR-based ReRAM devices not only demonstrated a significantly reduced off-current ($10^{-9}$ A) but also showed a markedly elevated on-off ratio ($\approx 10^5$) in comparison to previous reported low-dimensional perovskite and perovskite-derivative materials with similar device architectures.

## 2. Results and Discussion

The 'delta' form of $FAPbI_3$ represents a non-perovskite phase that is less explored compared to the more commonly studied 'alpha' form (perovskite phase). In this study, $\delta$-$FAPbI_3$ NPs were prepared via the hot injection method,[15] a proven technique for generating monodisperse colloidal NPs. This method involves rapidly injecting precursor solution into a hot solvent that contains other reagents and capping ligands, initiating fast homogeneous nucleation and subsequent diffusion-controlled growth. Owing to the size-focusing effect, larger NPs grow at a relatively lower rate than smaller NPs. As the reaction progresses, Ostwald ripening transpires, whereby larger NPs continue to grow, whereas smaller ones dissolve, driven by their higher chemical potential. By manipulating variables such as the temperature, surfactant concentration, and reaction time, the desired NPs can be obtained. This method, first introduced by Kovalenko et al. for synthesizing $CsPbX_3$ (X = Cl, Br, I) perovskite nanocrystals.[15b] In our experiment, formamidinium iodide was injected into a solution containing 1-octadecene, $PbI_2$, oleic acid, and oleylamine at 150 °C (see the experimental section in the Supporting Information for details). Notably, the reaction temperature was critical in obtaining $\delta$-FAPbI3 NPs; conducting the same reaction at 80 °C resulted in the $\alpha$-form.[10] The synthesized NPs were comprehensively characterized using various analytical techniques such as transmission electron microscopy (TEM), X-ray diffraction (XRD), and X-ray photoelectron spectroscopy (XPS), to determine their shape, size, crystallinity, and chemical composition. TEM images showed that the NPs were mainly spherical or hexagonal in shape (**Figure 1**a). The monodispersity of the NPs was quantitatively assessed through TEM, showing an average size of 16±2 nm (Figure S1, Supporting Information). The bright Fast Fourier Transform (FFT) image (Figure 1b) confirms the crystallinity and hexagonal structure of the NPs. The *d*-spacing values of 3.9511 nm and 2.3041 nm obtained from the FFT analysis confirmed the presence of the (002) and [3–12] planes, respectively. From the miller indices (*h*,*k*,*l*) and *d*-spacing values, the lattice parameters were found to be $a = b =$ 8.6639 Å and $c$ = 7.9019 Å, well supported by the reported single crystal data of bulk $\delta$-$FAPbI_3$ crystal.[13] The high-resolution TEM image shows an interplanar distance of 7.59 Å corresponding to the (010) plane (Figure 1c). The XRD spectrum of the NPs film (Figure 1d) matched that of the corresponding bulk material reported in the literature, confirming the phase purity of the synthesized NPs.[6d,16] This phase purity, along with the chemical composition, was further substantiated by XPS analysis (Figure S2, Supporting Information). Notably, our synthesis method resulted in a phase-pure material, in contrast to the bulk form of $\delta$-$FAPbI_3$ formed by the phase transition of $\alpha$-$FAPbI_3$.

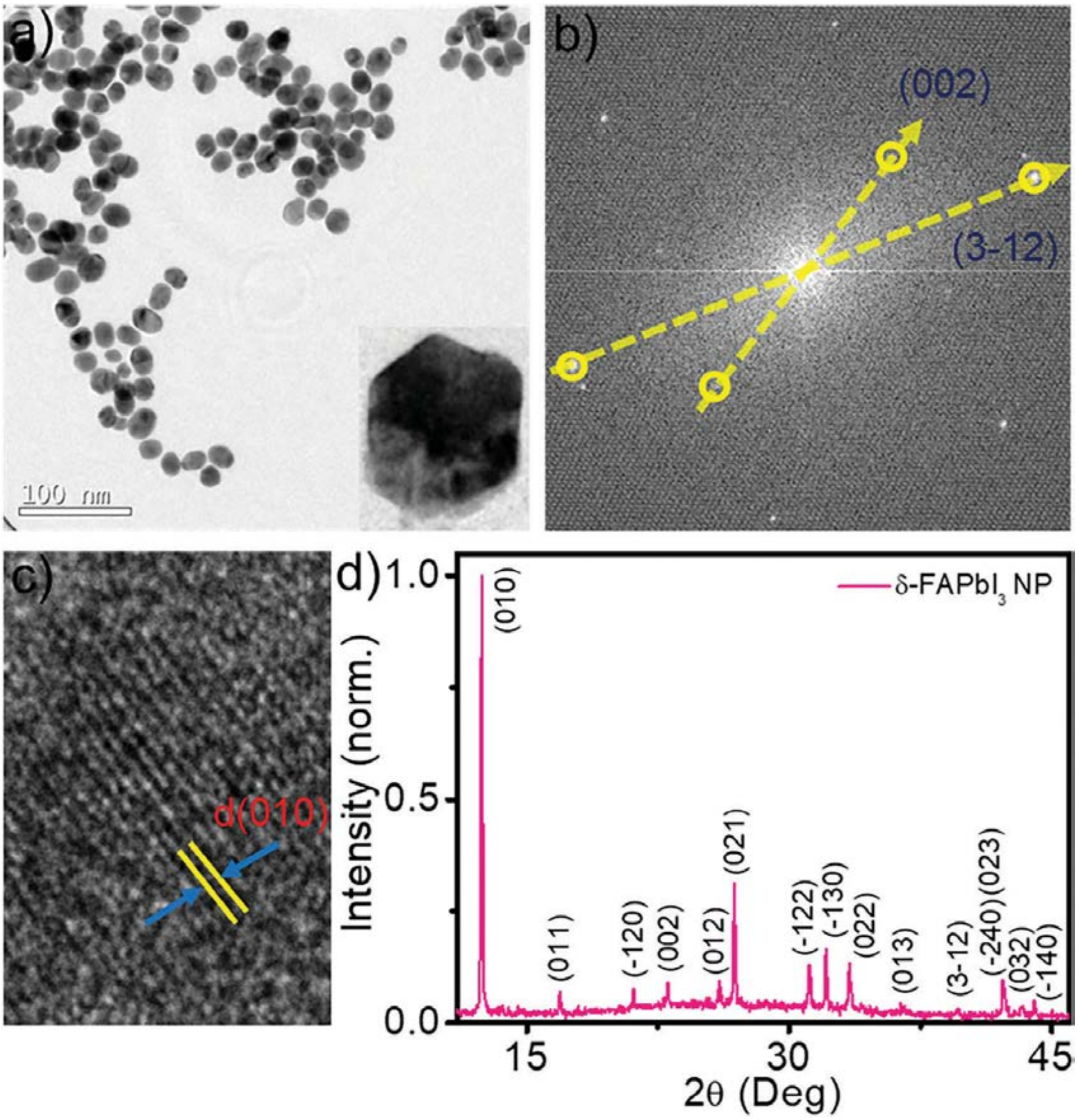

***Figure 1.*** *a) TEM image of $\delta$-$FAPbI_3$ NPs (average size of 16±2 nm). The inset shows the TEM image of a single NP exhibiting a hexagonal structure. b) Fast Fourier Transform image corresponding to the (002) and (3–12) planes. c) HR-TEM image of a typical $\delta$-$FAPbI_3$ NP. d) XRD spectrum of the NPs film.*

The NPs displayed a remarkable ability to self-assemble into NRs when transitioning from the colloidal to the film state. The self-assembly process was monitored by TEM at different stages of the growth reaction over the course of ≈8 h. The NPs were mostly spherical at low concentrations (**Figure 2**a), but with increasing concentration, they spontaneously orientated and organized along one direction, resulting in the formation of NRs (Figure 2b–d). This observation was further corroborated by scanning electron microscopy (SEM) analysis (Figure S3a–c, Supporting Information). The dynamic light scattering (DLS) profiles of both $\delta$-$FAPbI_3$ NPs and NRs are shown in Figure S4 (Supporting Information), which shed light on the change in size distribution due to the transformation of NPs to NRs. The topological SEM images showed that the NRs film was highly dense and pinhole-free, with the majority of them stacked parallel to the substrate (Figure 2e,f). Notably, the ligands in NPs of halide perovskites/perovskite-like materials significantly influence the growth kinetics, preferential orientation, and final morphology during the self-assembly process. In the present case, the spontaneous alignment and fusion of NPs into NRs are believed to be driven by interactions between the NPs following partial displacement of the capping ligands, which are known to be weakly bound and easily detachable.[17] Non-polar solvents such as toluene efficiently detaches these ligands, leading to direct surface contact between NPs and subsequently facilitating the formation of NRs.

Reports suggest that the binding energy of oleylammonium cations (OAm) onto the perovskite NP surface is much higher than that of oleate (OA) ligands.[18] This differential binding affinity results in the OA ligands detaching more readily from the nanocrystal surfaces during washing, yielding oriented attachment of NPs through facets with insufficient surface coverage of capping ligands. Furthermore, the inherent soft ionic lattice of the NPs facilitates their regrowth into a more energetically favorable NR configuration. Even without the purification step, the NPs undergo fusion, followed by a regrowth process during film formation. This phenomenon is driven by the elevated chemical potential of the NPs and strong Coulombic interactions between them in the film state. Consequently, individual NPs were indistinguishable in the SEM and TEM images of NPs at high concentrations or in the film state. The XRD profile of the NRs exactly matched that of the NPs (Figure 1d), indicating a consistent composition between them. A schematic diagram elucidating the self-assembly of $\delta$-$FAPbI_3$ NPs followed by growth into NRs, shown in Figure S5 (Supporting Information), shows the transformation process, offering a clearer insight into the proposed mechanism.

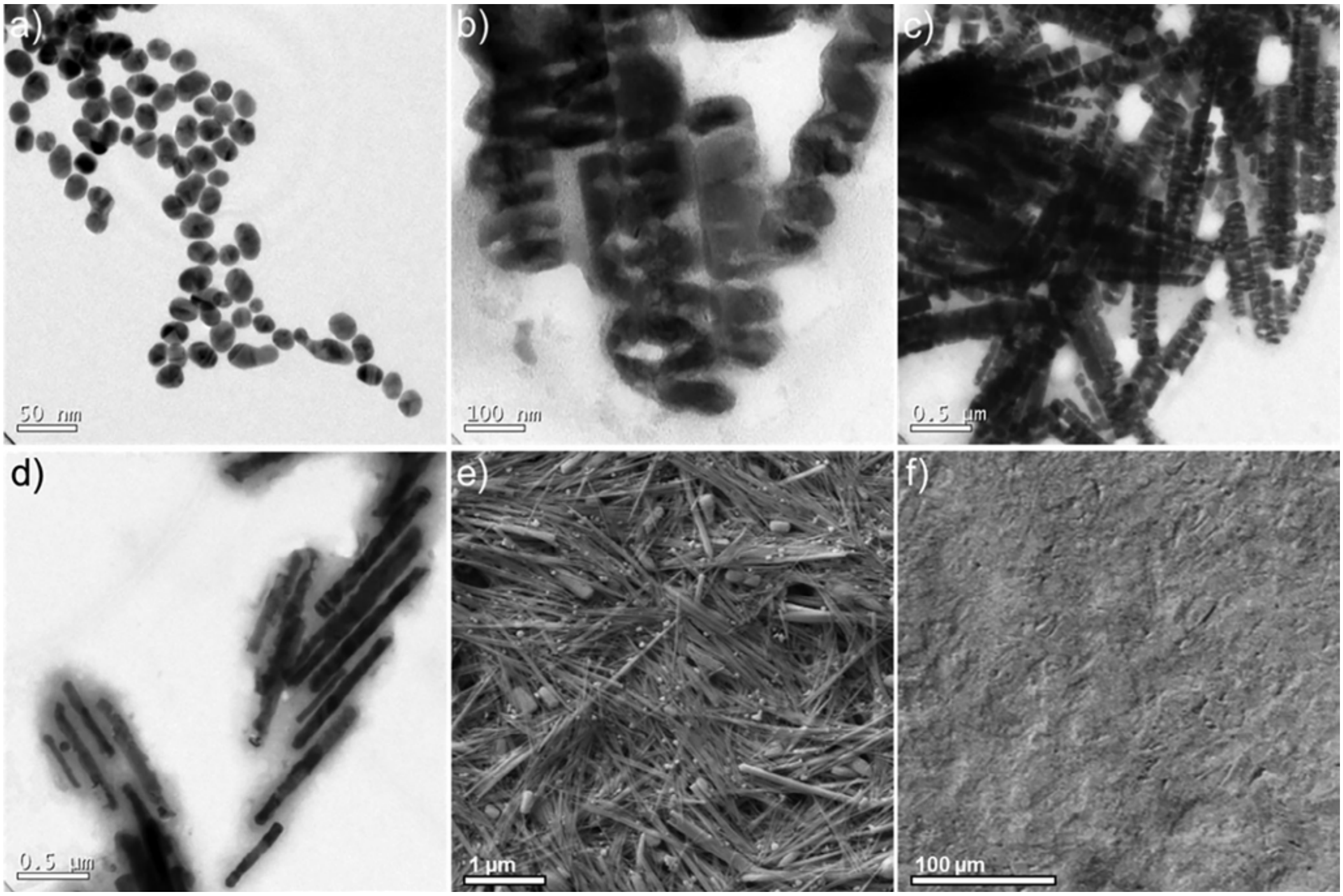

***Figure 2.*** *a–d) TEM images showing the self-assembly of δ-FAPbI3 NPs into NRs. The self-assembly process occurred for over 8 h. e,f) SEM image of the δ-FAPbI3 NRs film at two different magnifications.*

The absorption spectrum of the $\delta$-$FAPbI_3$ NPs dispersed in toluene exhibited absorbance up to 425 nm with an absorption maximum at 377 nm (Figure 3a; green), which was blue-shifted compared to that of $\alpha$-$FAPbI_3$ NPs ($\lambda_{max}$ = 768 nm) because of their low dimensionality and differences in the band structure, resulting from different lattice geometry (face-sharing octahedra for $\delta$-$FAPbI_3$ and corner-sharing octahedra for $\alpha$-$FAPbI_3$). The absorption spectrum of the $\delta$-$FAPbI_3$ NR film (Figure 3a; orange) was derived from the reflectance spectrum (Figure S6, Supporting Information) using the Kubelka-Munk function. It showed the onset of absorbance at 453 nm, with an absorption maximum at 422 nm. The bandgap energies, calculated using the Tauc equation and assuming an indirect bandgap, were found to be 3.00 eV for the NPs and 2.72 eV for the NRs. The absorption peak of the NR film exhibited a 45 nm red-shift relative to the NPs, corresponding to a bandgap energy difference of 0.11 eV. This red-shift is ascribed to the charge carrier delocalization along the NRs, leading to a less pronounced quantum confinement effect, due to the differences in size and dimensionality between the NPs and NRs.[19] Both the NPs in colloidal form and the NRs in film form displayed a pale yellow color (inset of Figure 3a). Consistent with theoretical and experimental findings [20], δ-$FAPbI_3$, being an indirect bandgap material, promotes non-radiative recombination, resulting in no detectable emission in either NP or NR forms.

In this study, the potential of the 1D NRs as the active layer in ReRAM devices was explored. The NRs, aligned parallel to the fluorine-doped tin oxide (FTO) substrate, were hypothesized to contribute to reducing the off-state current and enhancing the on/off ratio of the ReRAM devices. The device configuration used was a simple FTO/NR/Al structure, each with an area of 1 $mm^2$. A cross-sectional SEM image of the device revealed that the thicknesses of the NR film and aluminum electrode were ≈200 nm (**Figure 4**a). The current-voltage (*I*–*V*) characteristics of the device were evaluated by applying bias voltages in the range of −3 V to 3 V (Figure 4b).

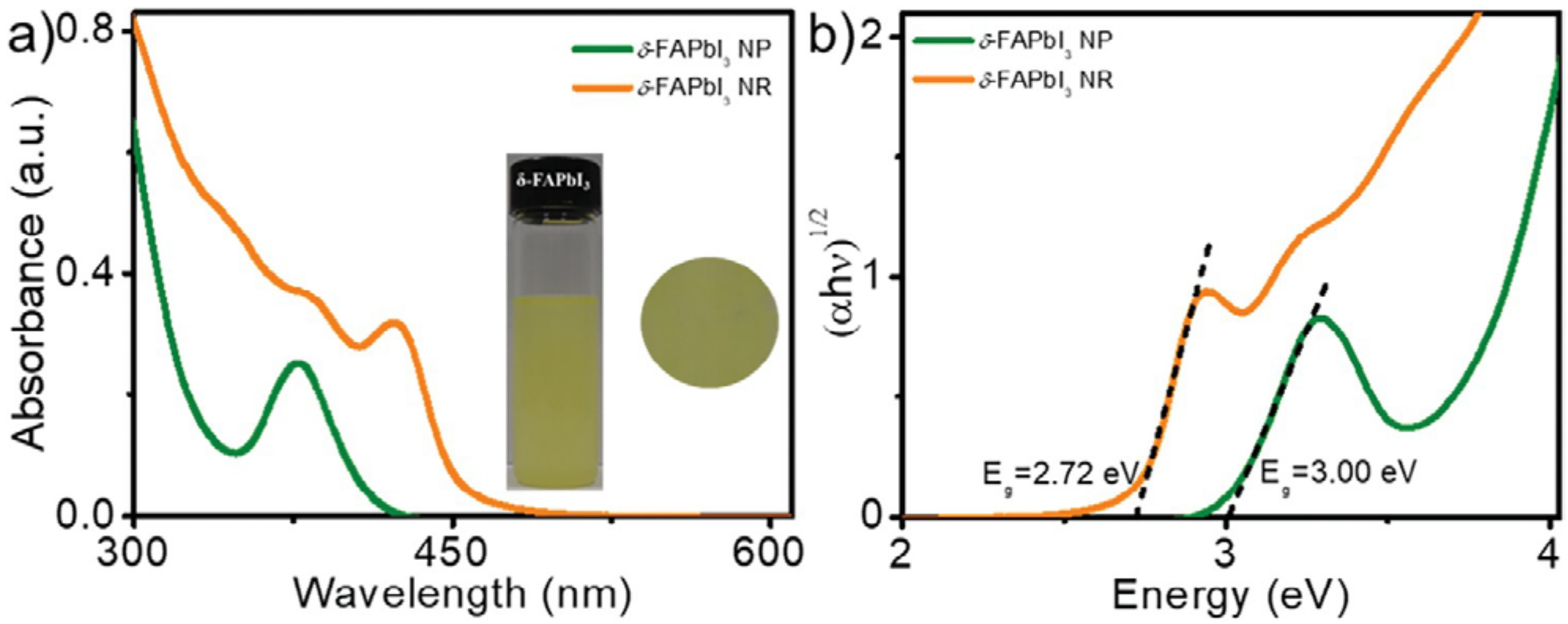

***Figure 3.*** *a) Absorption spectra of the δ-$FAPbI_3$ NPs dispersed in toluene (green) and NRs films on a quartz plate (orange). The corresponding photographs are shown in the insets of the figure. b) Determination of the corresponding band gap energies for NPs and NRs.*

The results indicate that the device transitions from an initially low resistance state (LRS) to a high resistance state (HRS) at 2.6 V during the RESET process and returns to the LRS at −3 V during the SET process. The repeatability of this bipolar resistive switching behavior was further validated through Weibull distributions of SET and RESET voltages measured for 25 devices (Figure S7, Supporting Information).

The cycling performance of the device was evaluated to demonstrate its repeatability in writing, erasing, and rewriting data for practical applications. This was conducted by applying consecutive DC voltage pulses and examining the electrical stability of the device at $V_{set} = -3$ V, $V_{reset} = 3$ V, and $V_{read} = 0.5$ V. The results shown in Figure 4c indicate that the devices exhibited stable resistive switching (data obtained for 250 cycles are shown in Figure S8, Supporting Information). The retention properties of the devices were evaluated at room temperature for 1000 s, with on/off states measured at 0.5 V, which indicated good stability of the devices in the programmed states with a remarkable on/off ratio of $10^5$ (Figure 4d). Notably, the on-state of these devices exhibited greater durability compared to devices based on $\alpha$-$FAPbI_3$ NPs,[21] possibly due to more stable conducting channels under a constant applied bias voltage. However, there was a fluctuation in the off current, which can be ascribed to charge trapping and detrapping events occurring in various trap states engendered by defects within the material. These defects, located at varying distances from the electrode, contribute to the charge carrier dynamics, resulting in the observed noise during off-state measurements.[22]

To further assess the long-term stability of the material and device, additional investigations were carried out. First, regarding the material stability over time, TEM and SEM analyses were carried out on the NRs after a one-year period (Figure S9, Supporting Information). The obtained images revealed that the material morphology was well-preserved, indicating enduring stability. The phase purity and absence of impurities were confirmed by XRD measurements (Figure S10, Supporting Information). Second, the resistive switching behavior of the devices was reexamined after one year. The *I–V* measurements (Figure S11, Supporting Information) demonstrated that the devices maintained their stability under ambient conditions. Additionally, a cross-sectional SEM image of the memory device after resistive switching and cycling (Figure S12, Supporting Information) revealed structural consistency comparable to the fresh sample, further affirming the material and device stability following resistive switching operations.

A detailed analysis of the current-voltage characteristics indicated a slightly asymmetric character of the current profiles under LRS conditions (**Figure 5**). The observed characteristics fit well with a leaky rectifying Schottky junction with one parallel and one series resistor. Assuming zero forward resistance of the ideal diode, the series resistance can be calculated to be 542 Ω, whereas the parallel resistance is 520 Ω. These values are too low to affect the performance of the device, particularly in the HRS state, where the resistance of the junction is larger by about five orders of magnitude (thus, the resistance of the barrier will contribute to the total resistance at the level of about ten ppm). These results indicate a very high-quality junction, in which both the filamentary processes and Schottky junction characteristics can be independently evaluated.

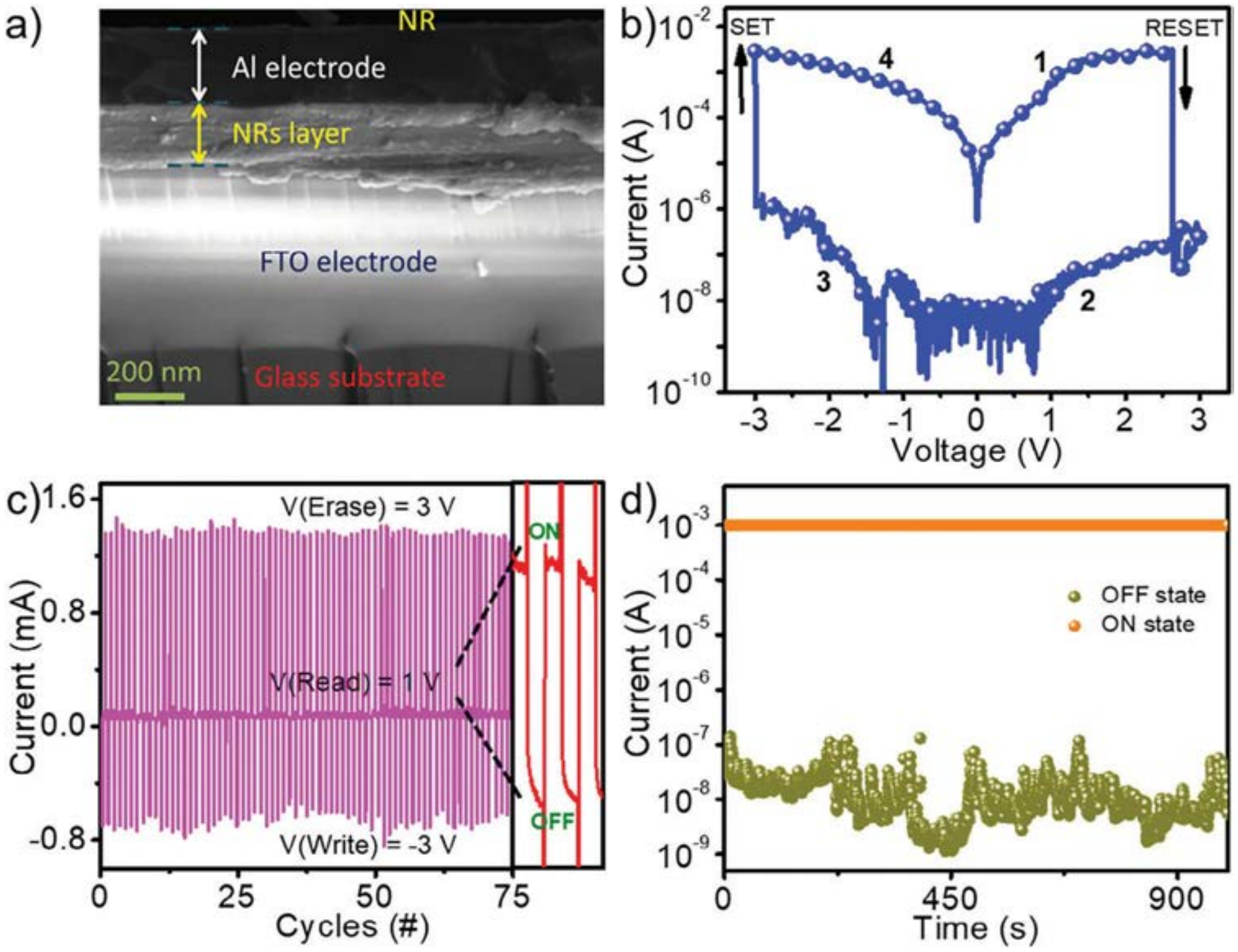

***Figure 4.*** *a) Cross-sectional SEM image of the FTO/NR/Al memory device. b) Typical bipolar resistive switching behavior of a memory device. c) Cycling tests of the devices measured for 75 cycles (*$V_{write}$ *= -3 V,* $V_{erase}$ *= 3 V,* $V_{read}$ *= 0.5 V). The cycling tests measured for 250 cycles are shown in the Supporting Information. d) Retention properties of the devices in the off and on states were measured at 0.5 V.*

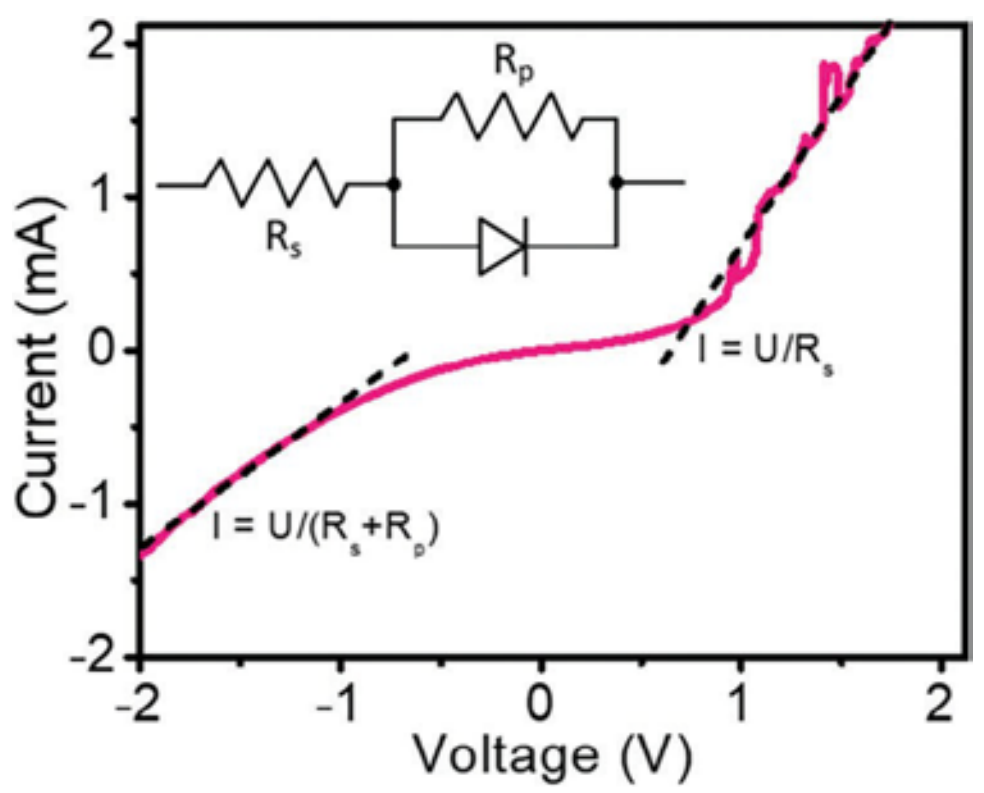

***Figure 5.*** *Current-voltage characteristics of the junction in the ON state. In the range of +1/+2 and −1/-2 V, the almost linear character allows for the determination of the junction series and parallel resistances.*

The device configuration used in the present study was FTO/NR/Al. With the work function of Al being 4.2 eV and that of FTO being 5 eV, a Schottky barrier is formed at the NRs/FTO interface, as evidenced by the asymmetric *I–V* profiles. This asymmetry indicates that the work function of the electrode material plays a significant role in the resistive switching behavior of the device. In theory, applying a reverse voltage could instigate similar effects as the movement of halide vacancies toward the bottom electrode and the consequent heat generated by current flow, in terms of disrupting the conductive channel. However, the reality might be more complex because of the Schottky barrier, which could impede the movement of charge carriers, depending on the polarity of the applied voltage. The Schottky barrier can foster rectifying behavior, which may exhibit different resistive switching characteristics under reverse-bias conditions. Furthermore, the contact materials and interface states can significantly influence the electric-field distribution, charge injection, and carrier transport within the device, thereby affecting the SET and RESET processes.

The resistive switching mechanism was studied in detail by examining a plot of log (*V*) against log (*I*) for the SET process (**Figure 6**a).[23] The slope of the HRS at a low electric field (represented by the orange line) is 1.2, corresponding to ohmic transport. The slope of 2.1 (green line) corresponds to the space charge limited current (SCLC), and a slope greater than 2 (red line) indicates tunneling and related mechanisms. The device was found to be in the TFLC region until the applied voltage was near −3 V, where the SET process occurred. At this point, the current suddenly increased, and the slope changed to ohmic conduction (S = 1.1), which confirmed the formation of conducting channels. The observation of ohmic, SCLC, and TFLC conduction suggests that the switching mechanism in the dielectric $\delta$FAPbI$_3$ NRs active layer is due to bulk-limited conduction rather than electrode-limited conduction, which is consistent with high on/off ratio, and reverse/forward current ratio at LRS close to unity.[23c,24] Since FAPbI$_3$ has two distinct phases, and there may be the possibility that the memory effect originates from a phase change between these phases.[25] To investigate this possibility, the XPS profile of the material was analyzed in the high resistance state (HRS; Figure S13, Supporting Information). The results indicated that the XPS profile remained unchanged compared with that of the low-resistance state (LRS), excluding the possibility of a phase-change-based resistive switching mechanism. It was further supported by additional XRD measurements (Figure S14, Supporting Information) on the ReRAM devices both before and after resistive switching to further elucidate any potential phase changes. XRD analysis revealed identical peaks in the spectra before and after resistive switching, thus confirming the absence of a transition to the alpha phase during this process. As previously mentioned, the device initially exhibited a LRS, suggesting the existence of conducting channels within the active layer even before the application of any external bias voltage. Considering the high migration energy associated with the movement of organic cations and $Pb^{2+}$ ions, the observed conduction mechanism is likely driven by halide ion vacancies ($V_I$). The low migration energy for $V_I$ is a reflection of the ionic character and soft ionic lattices inherent to perovskite/perovskite-derived materials. In these materials, halide ion migration can occur along the edges of the octahedral $[PbI_6]^{4-}$ structure. Moreover, the $V_I$ situated in both axial and equatorial sites are found to be isoenergetic, contributing to the low migration energy when compared to other point defects present in perovskite materials.[26] The process of NRs formation from NPs via orientational attachment followed by fusion/regrowth does indeed lead to a reduction in $V_I$ concentration due to the low surface-to-volume ratio of NRs. The $V_I$ present in the facets of the NPs and the post-detachment of weakly binding capping ligands facilitate this orientational attachment. Although the other facets covered with stronger capping ligands do not participate in NR formation, complete passivation is impeded by the repulsion energy between the long alkyl chain capping ligands. Therefore, a substantial number of $V_I$ remain in the structure, albeit at a reduced concentration compared to that of monodispersed NPs. This phenomenon has been previously reported in perovskite-based memory devices, where the presence of $V_I$ was identified as the cause of the initial LRS.[10,27] Various literature reports reinforce this hypothesis, showing resistive switching mechanisms in perovskite-based memory devices driven by the formation and rupture of conducting filaments formed by $V_I$.[6a,28]

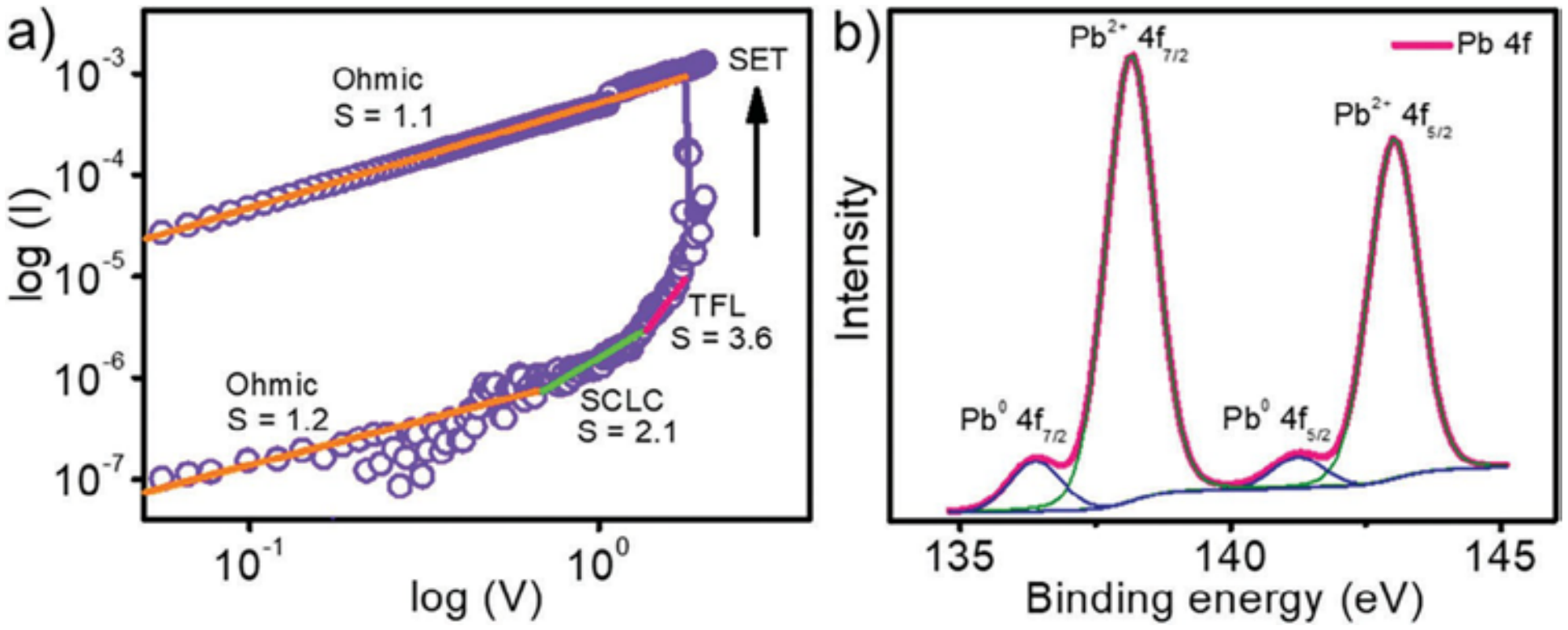

***Figure 6.*** *a) Logarithmic I–V plot of the memory device during SET process. b) XPS spectrum of Pb 4f corresponding to the δ-FAPbI3 NR film coated on the FTO substrate.*

To elucidate the cause of the initial LRS in the devices, XPS analysis was performed on the NR film. The XPS spectrum of Pb 4f (Figure 6b) revealed the presence of both Pb2+ (at 138.1 eV and 143.0 eV) corresponding to the ions housed within the inorganic octahedra of [PbI6]4− in the material and low-valent Pb0 (at 136.2 eV and 141.1 eV) peaks. The metallic Pb0 peak can be seen as an indirect yet clear identification of the VI in the active layer. It is believed that during the purification of the δ-FAPbI3 NPs, an increased number of VI may have been generated because of the partial detachment of the capping ligands.[18,29] These VI facilitate a redox reaction that reduces the interstitial $Pb^{2+}$ ($Pb^{\bullet\bullet}$) to low-valent lead species ($Pb^0$).[30] The reaction can be represented as follows ($V_I$ are compensated by electrons):

$$2e' (V_I) + Pb_i^{\bullet\bullet} \rightarrow Pb^0 \qquad (1)$$

The identification of metallic lead at various thicknesses of the NRs film using in-depth XPS analysis (Figure S15, Supporting Information) confirmed the presence of $V_I$ throughout the film. These $V_I$ were adequate for the formation of conducting filaments, thereby placing the device in an initial LRS without requiring a forming or set process. This phenomenon aligns with observations made in previous studies on $MA_3Sb_2Br_9$ and $GdO_3$ based ReRAMs.[31] Though zero-valent Pb is present in the film, the formation and rupturing of the conducting channels are exclusively caused by $V_I$ because Pb is difficult to move. During the RESET process, it was observed that the self-formed conducting filaments easily ruptured owing to the movement of $V_I$ to the bottom electrode and the heat generated by the current flow.

The *I–V* measurement in orthogonal direction was conducted to provide a more comprehensive understanding of the charge transport behavior between the NRs (Figure S16, Supporting Information). This measurement exhibit a current in the picoampere range, indicating that conduction along the NRs is significantly subdued. The anisotropy in the NRs was elucidated through the exploration of charge-carrier migration pathways. The idea of calculating the migration barrier of VI in a few unit cells of delta and alpha phases is insightful; however, implementing such a computation presents notable challenges due to the high complexity arising from the capping ligands and their interaction with the VI. Nevertheless, the theoretical evidence reported in literature offers some insights. For instance, density functional theory calculations reported by Yang et al. supports the anisotropic migration behavior of $V_I$ in $\delta$-FAPbI3.[6d] This investigation revealed two distinct migration pathways: [010] and [123]. Notably, the [010] direction is aligned parallel to the "*ab*" plane, whereas the [123] direction is slightly tilted from the *c* axis. A critical facet of these findings is the substantial migration barrier encountered by VI along the [010] direction, with a value of 0.9 eV, attributed to the presence of formamidinium cations. On the other hand, a lower migration barrier (0.48 eV) was observed along the [123] direction. The *I–V* measurements conducted in this study reflect a picoampere (pA) current in the orthogonal direction, which further substantiates the notion that the "*ab*" plane is oriented perpendicular to the electrodes. The orientation of the NRs along the FTO substrate further suggested that there were fewer halide vacancies in the orthogonal direction. Consequently, it was established that the NRs exhibit poor conduction along their elongated axes. A schematic representation of the two possible pathways of charge-carrier migration is shown in Figure S17 (Supporting Information). Despite a low energy barrier for vacancy migration, this material shows relatively high switching potential (2.6–3.0 V). This is related to the presence of long-chain aliphatic capping ligands at the surface of nanowires. These organic layers act as insulators and reduce the electric fields within the NRs, which effectively increases the activation energy required for vacancy migration.

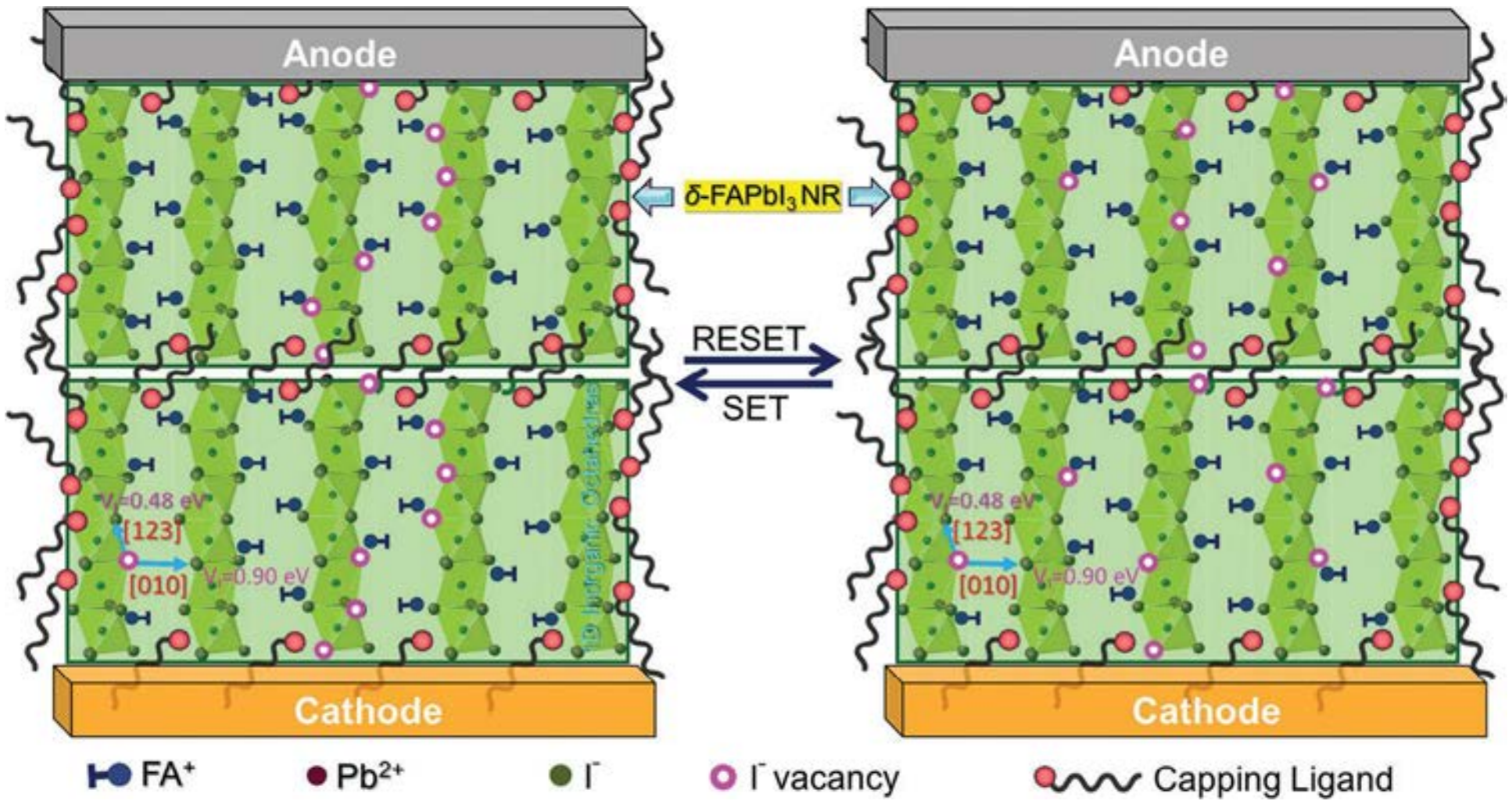

***Figure 7.*** *Schematic diagram illustrating SET and RESET processes of the memory device.*

The unique characteristics of NRs in the vertical direction could be attributed to two major factors: the Schottky barriers between the rods and halide ion migration within the crystal crosssectional area. TEM and SEM analyses confirmed that the NRs were oriented parallel to the FTO substrate. A multitude of unit cells are included in the vertical dimension of an NR. These 1D inorganic octahedra can be aligned either parallel or perpendicular to the NR axis. The observed poor conductance along the direction orthogonal to the NR axis indicates that these 1D inorganic octahedra are aligned perpendicularly relative to the FTO substrate. This alignment is further corroborated by the initial LRS of the device, which indicates high conductance. This implies that the $V_I$ are oriented perpendicular to the electrodes, establishing a conducive path for electrical conduction. Lastly, the presence of capping ligands on the NRs contributes to formation of energy barriers between the parallelly arranged rods.[32] The robustness of the NRs' structure, which retains its size even after a year, implies a very low interaction energy between parallel NRs. This suggests that the capping ligands create a substantial barrier hindering the hopping of halide vacancies, which is likely to contribute to the observed lower off-current. This arrangement minimizes the interaction energy among $V_I$ in adjacent NRs, thus facilitating an easier rupture of the conducting filaments and attenuating the hopping of charge carriers between the electrodes. The interplay between the Schottky barriers between the NRs and the migration of $V_I$ within the cross-sectional area of the crystal is pivotal in explaining the unique electrical characteristics of the NRs.

It has been observed that the on/off ratio of the present devices ($10^5$) is significantly higher in comparison to that of our previously reported $\alpha$-$FAPbI_3$ NPs[10] based memory devices (device 26, **Table 1**). This was attributed to the low off-state current of the former. The comparison between the on-currents of $\delta$-$FAPbI_3$ NRs and $\alpha$-$FAPbI_3$ NPs suggests that halide vacancies play a similar role in conductive filament formation. However, the different structures of these two materials, with $\delta$-$FAPbI_3$ being 1D and $\alpha$ $FAPbI_3$ being 3D (Figure S18, Supporting Information), significantly impact the dynamics of charge carrier hopping and subsequently, the off-state current levels. The fusion of perovskite NPs into NRs could lead to a reduction in the halide vacancy concentration owing to a decrease in the surface-to-volume ratio, thereby reducing the overall defect concentration. Although both materials exhibit similar on-currents, the structural disparity results in a thicker and stronger conducting filament in $\alpha$-$FAPbI_3$ NPs because of the 3D network of inorganic octahedra compared to the 1D arrangement in $\delta$-$FAPbI_3$ NRs. This structural difference directly affects the ease of charge-carrier hopping between halide vacancies, leading to a higher off-current in the $\alpha$-$FAPbI_3$ NPs. Moreover, the interaction energy between halide vacancies is reported to be higher within 3D structures compared to 1D structures [6d] which, in conjunction with thicker conducting filaments in $\alpha$-$FAPbI_3$ NPs, might hinder the complete rupture of conducting filaments during the RESET process, thus contributing to a higher off-current. These observations delineate the arrangement of the NRs, 1D inorganic octahedra, and halide vacancy filaments, as depicted in **Figure 7**a. Throughout the RESET process (Figure 7b), it was determined that the autonomously formed conducting filaments were easily disrupted owing to the mobilization of halide vacancies toward the bottom electrode coupled with the thermal energy engendered by the current flow.

Finally, the RS properties of the current device were compared with those of previously reported low-dimensional bulk and three-dimensional nanostructured perovskite-based ReRAM devices with similar device architectures (Table 1). A high on/off ratio with a low on-current (LRS) and off-current (HRS) are important parameters that determine the performance of memory devices. The off current in perovskite devices is associated with the vacancy-mediated migration of halide ions. Therefore, reducing the concentration of halide ion vacancies is important for suppressing halide ion migration under an applied electric field to reduce the off-current. Most importantly, the suppression of halide ion migration also improved stability. The LRS is related to the number and density of conducting filaments formed by the halide vacancies. There should be a trade-off to control the HRS and LRS to achieve a high on–off ratio.

The literature analysis indicates that the attempt to reduce the off-current demands compromises other parameters such as the on-off ratio and the thickness of the films. On the other hand, maintaining a very low off-current is very important in memristive devices to increase their energetic efficiency. In perovskite-based devices, the off current is determined by the halide ion vacancy concentration and distribution. These vacancies promote the migration of halide ions to the corresponding electrode under an electric field, leading to degradation of these materials. In the case of low-dimensional bulk perovskites,the off-state current can be reduced by preparing a single crystal (device 1), increasing the grain size to reduce halide vacancies (device 4), and increasing the active layer thickness at the micrometer scale (devices 6 and 7). Although the devices have a low off current, they show a very low on-current, leading to a low on/off ratio due to the low surface area in bulk perovskite and a limited number of halide vacancies in single crystals. To improve the oncurrent and thus the on/off ratio, an electroactive Ag electrode was used. However, silver ions can interact strongly with halide ions to form AgX,[53] which leads to the degradation of the perovskite active layer. In some cases, the external light source has been used to increase the on current (device 13), which could also promote perovskite degradation.[36]

Devices fabricated using colloidal NPs possess a high off-state current (devices 15 and 16) because of the large number of surface defects. Template-assisted synthesis of nanostructured perovskite requires ultra-high thickness (10-20 μm) to obtain a low off-current (devices 22–25). An advanced evaporation technique was used to develop the template layer, and an additional layer ($Al_2O_3$; device 24) was used for the long-term stability of the device. Finally, the RS properties of the current device were compared with $\alpha$-$FAPbI_3$ based memory device (device 26). Both devices possess almost similar on-state current ($10^{-3}$ A) and active layer thickness (200 nm), which indicates that the halide vacancies involved in the conducting filaments are almost similar. However, $\delta$-$FAPbI_3$ NR-based devices have a low off current (four orders less), which may stem from their low dimensionality, reduced number of defects due to the fusion of NRs, thickness of conducting filaments, interaction energy between halide vacancies, and parallel orientation. Our device showed a higher on/off ratio with lower onand off-state currents than those previously reported.

The material explored in this study shows promising potential for further miniaturization of memristive devices. This can be achieved by reducing the footprint and layer thickness of individual memristors.

However, such modifications pose potential challenges, including possible impairments to the electrical properties of nanowires. This is due to the quantum size effect and the increased role of capping ligands in the overall electrical transport phenomena. Looking forward, another promising avenue in the development of NR-based memristors is the improved control over the alignment of individual NRs into parallelly oriented layers. While this approach has the potential to enhance device performance, it is important to note that it would also significantly increase the complexity of device fabrication. Such advancements will require careful balancing of performance benefits against the practicality of manufacturing processes.

***Table 1.*** *Comparison of resistive switching properties of δ-$FAPbI_3$ NRs based memory devices with other perovskite and perovskite-derivative materials with similar device architectures.*

| | Device structure | $[V_{set}/V_{reset}]$ | Endurance [cycles] | Retention [time, s] | $I_{off}$ [A] | on/off ratio | Year | Ref. |
|---|---|---|---|---|---|---|---|---|
| 1 | FTO/Graphene/2D $(PEA)_2PbBr_4$ / Au | +2.8/−1.0 | 100 | $10^3$ | $10^{-13}$ | 10 | 2017 | [33] |
| 2 | ITO/2D $MA_2PbI_2(SCN)_2$ /Al | -1.59/−3.20 | - | $10^4$ | $10^{-9}$ | $10^4$ | 2018 | [34] |
| 3 | ITO/0D $MA_3Bi_2I_9$ /Au | +1.6/−0.6 | 300 | $10^4$ | $10^{-5}$ | $10^2$ | 2018 | [11a] |
| 4 | ITO/2D $BA_2PbBr_4$ /Au | | | | | | 2019 | [11b] |
| | Grain size = 180 nm | +3.0/−3.0 | 60 | $10^3$ | $10^{-4}$ | 5 | | |
| | Grain size = 1 μm | +3.0/−3.0 | 60 | $10^3$ | $10^{-5}$ | 40 | | |
| | Grain size = 30 μm | +3.0/−3.0 | 60 | $10^3$ | $10^{-7}$ | $2.4\times10^3$ | | |
| 5 | ITO/2D $CsBi_3I_{10}$ /Al | -1.7/+0.9 | 150 | $10^4$ | $10^{-5}$ | $10^3$ | 2019 | [35] |
| 6 | FTO/0D $Cs_3Bi_2I_9$ /Ag($AgO_x$) | -0.12/+1.0 | 250 | $10^6$ | $5\times10^{-8}$ | $10^6$ | 2020 | [36] |
| 7 | FTO/2D $CsBi_3I_{10}$ / Ag($AgO_x$) | -0.14/+0.55 | 250 | $10^6$ | $2\times10^{-8}$ | $10^5$ | 2020 | [36] |
| 8 | ITO/3D/2D | +0.79/−0.77 | 300 | $10^4$ | $10^{-6}$ | $10^3$ | 2020 | [8] |
| 9 | ITO/0D $MA_3Bi_2I_9$ /Cu | +1.0/−6.4 | 1730 | $3.0\times10^5$ | $10^{-7}$ | $10^4$ | 2021 | [37] |
| 10 | ITO/0D $Cs_3Bi_2Br_9$/Ag | -0.5/−1.0 | 3200 | $10^3$ | $10^{-5}$ | $10^2$ | 2021 | [38] |
| 11 | ITO/1D (NH = CINH$_3$)$_3$PbI$_5$/Au | +0.2/−2.1 | 200 | $10^4$ | $10^{-6}$ | $10^3$ | 2021 | [39] |
| 12 | ITO-PET/2D $CsPb_2Br_5$/Al | +2.34/−2.04 | 100 | $10^3$ | $10^{-8}$ | $10^2$ | 2021 | [40] |
| 13 | ITO/2D $Cs_3Bi_2Br_9$/Al | -0.45/+2.2 | 1000 | $10^4$ | $10^{-7}$ | $10^5$ | 2022 | [41] |
| 14 | ITO/2D $Cs_3Bi_2Br_9$/Ag | +0.44/−0.38 | 2000 | $10^4$ | $10^{-4}$ | $10^2$ | 2022 | [42] |
| 15 | FTO/ | +0.7/−0.5 | 250 | $10^3$ | $10^{-4}$ | 500 | 2016 | [9] |
| 16 | $MAPbBr_{1.97}Cl_{1.03}$/Ag | +1.0/−1.5 | - | $2.5\times10^4$ | $10^{-5}$ | 20 | 2016 | [43] |
| 17 | PET/ITO/$Cs_3Bi_2I_9$/Au | +0.3/−0.5 | 1000 | $10^4$ | $10^{-4}$ | $10^3$ | 2017 | [44] |
| 18 | Ag/$MAPbI_3$/Ag | +1.0/−0.4 | 75 | 750 | $10^{-9}$ | 7.5 | 2018 | [45] |
| 19 | Ag/$MAPbI_3$/Ag | - | - | $10^4$ | - | 8 | 2019 | [46] |
| 20 | ITO/$CsPbBr_3$/Au | -2.4/+1.55 | - | $10^3$ | $10^{-4}$ | $10^4$ | 2019 | [47] |
| 21 | ITO/$CsPbBr_3$/Au | +1.5/WORM | - | $10^3$ | $10^{-4}$ | $10^4$ | 2019 | [48] |
| 22 | Au/$Cs_3Sb_2Br_9$/Au | | | | | | 2020 | [49] |
| | Thickness = 5 μm | +2.1/−4.5 | - | $2\times10^4$ | | 106 | | |
| | Thickness = 10 μm | +2.6/volatile | 200 | - | $10^{-5}$ | $10^3$ | | |
| 23 | Al/$MAPbI_3$/Al. | +1.66/−0.47 | 500 | $10^4$ | $10^{-8}$ | $10^6$ | 2020 | [50] |
| 24 | Ag/$MAPbI_3$/Au | +2.4/−2.2 | $6\times10^6$ | 2 years | $10^{-5}$ | $10^7$ | 2021 | [51] |
| 25 | Ag/$MAPbCI_3$/Al | +5.0/−3.0 | $1.07\times10^6$ | $7\times10^9$ | $10^{-8}$ | $10^7$ | 2021 | [52] |
| 26 | FTO/$\alpha$-$FAPbI_3$ | -2.4/+2.3 | 1200 | 1000 | $5\times10^{-4}$ | 20 | 2021 | [10] |

| | | | | | | | |
|---|---|---|---|---|---|---|---|
| 27 | PNC/Al FTO/$\delta$-$FAPbI_3$ PNC/Al | -3/2.6 | 250 | 1000 | $5\times10^{-8}$ | $>10^5$ | This Work |

## 3. Conclusion

In summary, novel yellow-phase hexagonal $\delta$-$FAPbI_3$ perovskitederivative NPs were synthesized via a hot injection method. A key finding was that halide ion vacancies on the NP surface facilitated the fusion of individual NPs into one-dimensional NRs. These NRs exhibited excellent colloidal and environmental stability. Memory devices fabricated using these NRs displayed outstanding resistive switching properties, with a high on/off ratio of $10^5$, which is four orders of magnitude higher than their corresponding alpha form. This high on/off ratio was attributed to the low current in the HRS, which was, in turn, caused by the low dimensionality and parallel orientation of the NRs toward the FTO substrate. This study is the first to demonstrate the significance of the self-assembly of perovskite-derivative NPs into different morphologies and the importance of dimensionality and alignment on the substrate in terms of memristor properties. The performance of $\delta$-$FAPbI_3$ NR-based ReRAMs can be enhanced further through device engineering. Additionally, owing to their low dimensionality and environmental stability, $\delta$-$FAPbI_3$ NRs may be promising candidates for other applications such as capacitors and catalysts.

**Acknowledgements**

C.V. thank the Department of Science and Technology (DST) for funding (CRG/2020/002756) and the Indo-Poland Project (DST/INT/POL/P43/2020). K.S. was partially supported by Polish National Science Centre within the OPUS programme (grant agreement No. UMO2020/37/B/ST5/00663, by the NAWA programme (contract No. PPN/BIN/2019/1/00100) and by the “Excellence initiative–research university” programme for the AGH University of Krakow. C.M. and A.A. are grateful to CSIR and UGC for their research fellowships.

**Conflict of Interest**

The authors declare no conflict of interest.

**Data Availability Statement**

The data that support the findings of this study are available from the corresponding author upon reasonable request.

**Keywords: formamidinium lead iodide, nanomaterials, non-volatile memory, resistive switching, self-assembly**

Small

# Supporting Information

**1.** Experimental Section

*Chemicals*: Lead iodide (99.999%, Sigma Aldrich), Formamidinium iodide (98%, TCI Chemicals), Oleylamine (70%, Sigma Aldrich), Oleic acid (99%, Alfa Aesar), 1Octadecene (90%, TCI Chemicals) and Toluene (99.8%, Merck) were used without any further purification.

*Materials and methods:* $^{1}H$ NMR spectra were recorded on a 500 MHz Bruker Avance DPX spectrometer. TEM measurements were carried out using an FEI-TECNAI T30 with EDAX at an accelerating voltage of 300 kV. A UV-Vis spectrophotometer (Shimadzu UV-2600) was used to record the UV-Vis absorption spectra of the perovskite materials in the solution and solid states. The optical properties of the solution were measured using a quartz cuvette with a 1 cm path length, and the solution was drop-cast onto a quartz plate dried under humid conditions for the film state measurements. The X-ray diffraction data were collected on a Rigaku AFC-12 Saturn 724+ CCD diffractometer equipped with a graphite-monochromatic Mo Kα radiation source ($\lambda$ = 0.71073 Å) and a Rigaku X-Stream at 293 K. Samples for analysis were prepared by drop-casting the precursor solution on a washed glass plate at the required concentrations and dried under ambient conditions. The surface chemical composition and oxidation states of the constituent elements present in $\delta$-FAPbI3 PNCs were analyzed by X-ray Photoelectron Spectroscopy (XPS) using a PHI 5000 Versa Probe II (ULVAC-PHI Inc., USA) equipped with micro focused (100-200 µm, 15 kV) monochromatic Al-Kα X-Ray source (hν=1486.6 eV). First survey scans were acquired on the samples and for the major detected elements, highresolution spectra were recorded. These spectra were used for estimating elemental compositions (%At.) and chemical state assignments by a curve fitting software (Multipak). Survey scans were recorded with X-ray source power of 23.7 W and pass energy of 187.85 eV.

XPS depth profiles are built by successively alternating 100 s sputtering with 2 keV $Ar^{+}$ beam focusing in the area of 2x2 mm active layer to obtain profile data up to 200 nm depth and XPS analysis performed with a monochromatic Al-kα X-ray beam with a spot size of 200 µm.

Topological and cross-sectional scanning electron microscopy (SEM) imaging of the device was performed by subjecting the device to a thin gold coating using a JEOL JFC-1200 fine coater. The probe side was inserted into a JEOL JSM-5600 LV scanning electron microscope for imaging purposes.

*Synthesis of δ-$FAPbI_3$ nanoparticles (NPs):* 5 mL of 1-octadecene was heated to 150 °C under vacuum in an oil bath, and 87 mg of $PbI_2$ (0.187 mmol) was added to 1-octadecene. Then, 0.8 mL of oleic acid and 0.35 mL of oleylamine were added to 1-octadecene under inert atmosphere, and the mixture was heated until complete solubilization of lead iodide. Next, formamidinium iodide (58.3 mg, 0.339 mmol) dissolved in 100 µL of DMF and 0.8 mL of oleic acid was injected, resulting in NP precipitation. Finally, the obtained NPs were separated by centrifugation at 12000 rpm for 10 min and washed twice with a mixture of 3:1 Toluene: ACN.

*Self-assembly studies:* 20 mg of freshly prepared and purified NPs in powder form was initially dispersed in 2 mL of toluene. Subsequently, the dispersion was diluted to the desired concentrations (5, 10, and 20 times) to achieve varying concentrations. The diluted solutions were then drop-cast onto the substrates for further analysis.

**2.** Size distribution curve

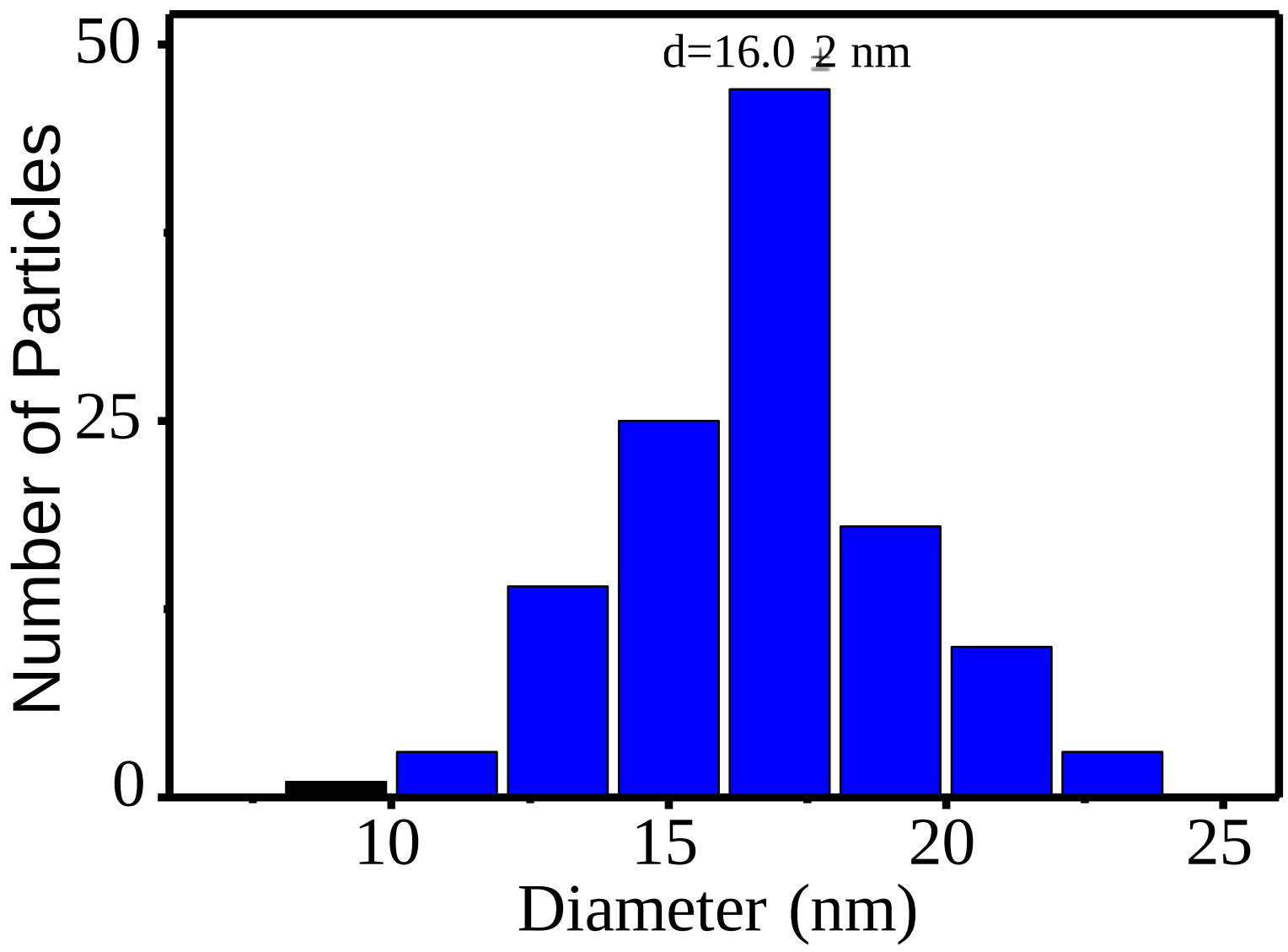

**Figure S1.** The size distribution curve of $\delta$-$FAPbI_3$ NPs obtained from the TEM image (Figure 1a) using GATAN Digital Micrograph software (Ver. 2.31.734.0). The average diameter of NPs was 16±2 nm.

**3.** XPS analysis

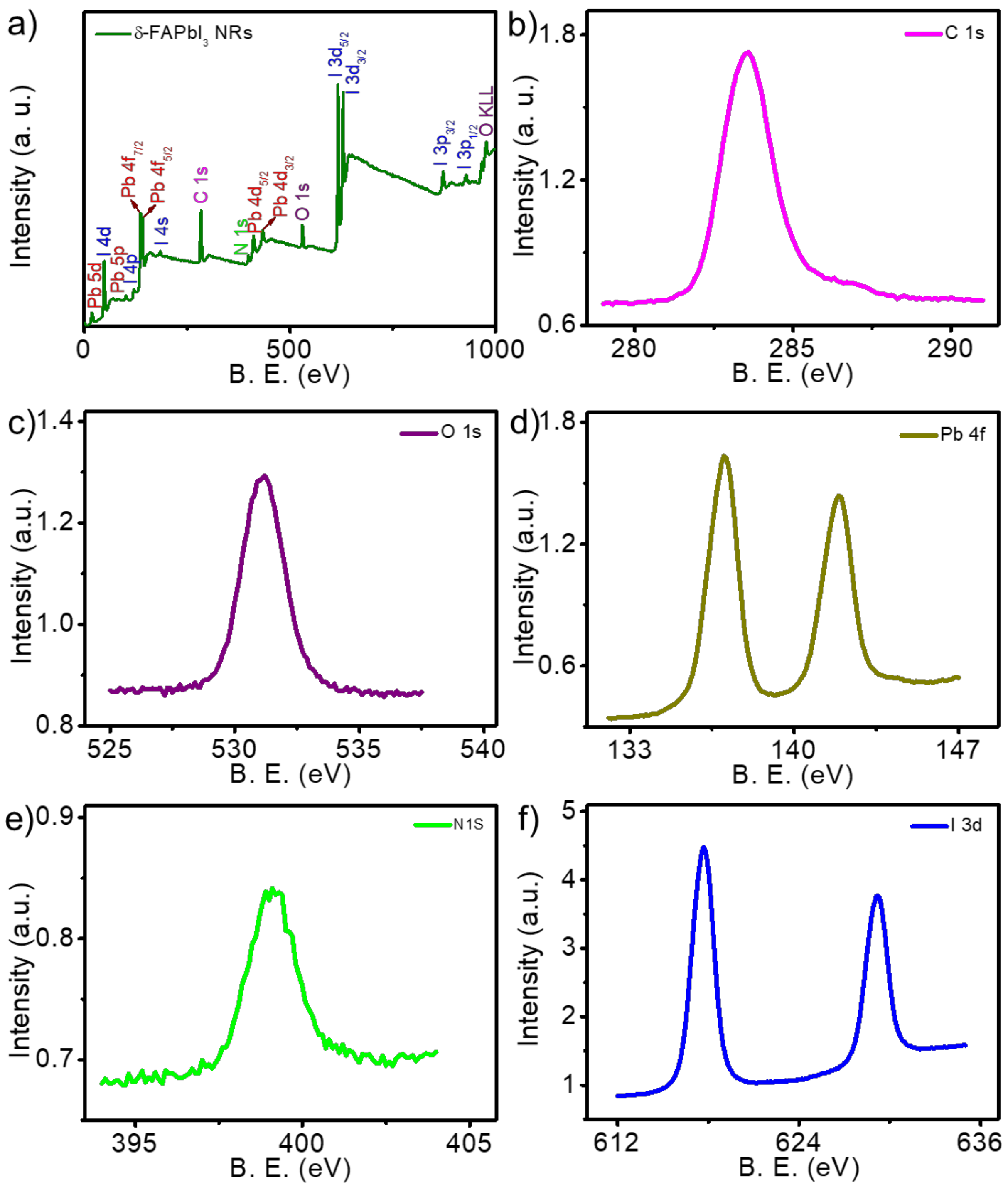

**Figure S2.** a) XPS survey spectrum of $\delta$-$FAPbI_3$ NPs. XPS profiles of NPs corresponding to b) C 1s, c) O 1s, d) Pb 4f, e) N 1s, and f) I 3d. The absence of additional peaks confirms the phase purity of the nanocrystals.

## **4.** SEM analysis

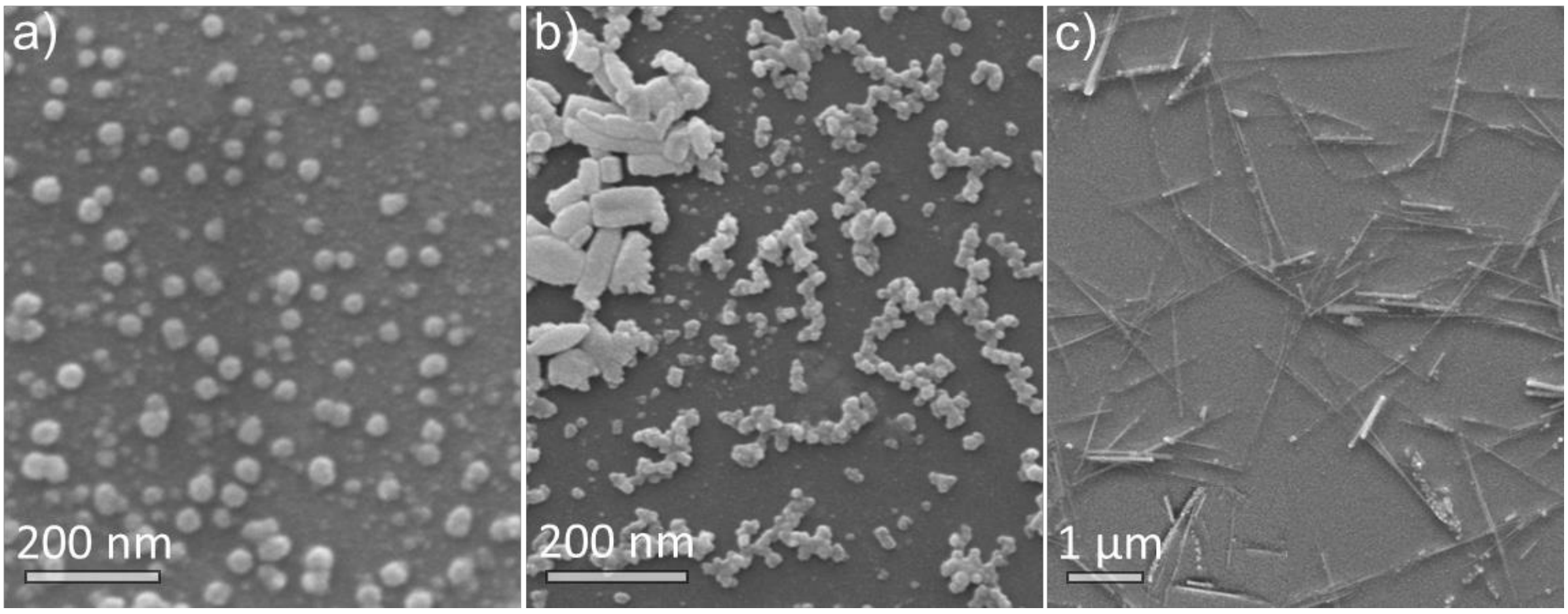

**Figure S3.** SEM images illustrating the self-assembly of $\delta$-$FAPbI_3$ NPs followed by fusion into NRs over 8 h. Each image represents a different stage in the process.

## **5.** DLS analysis

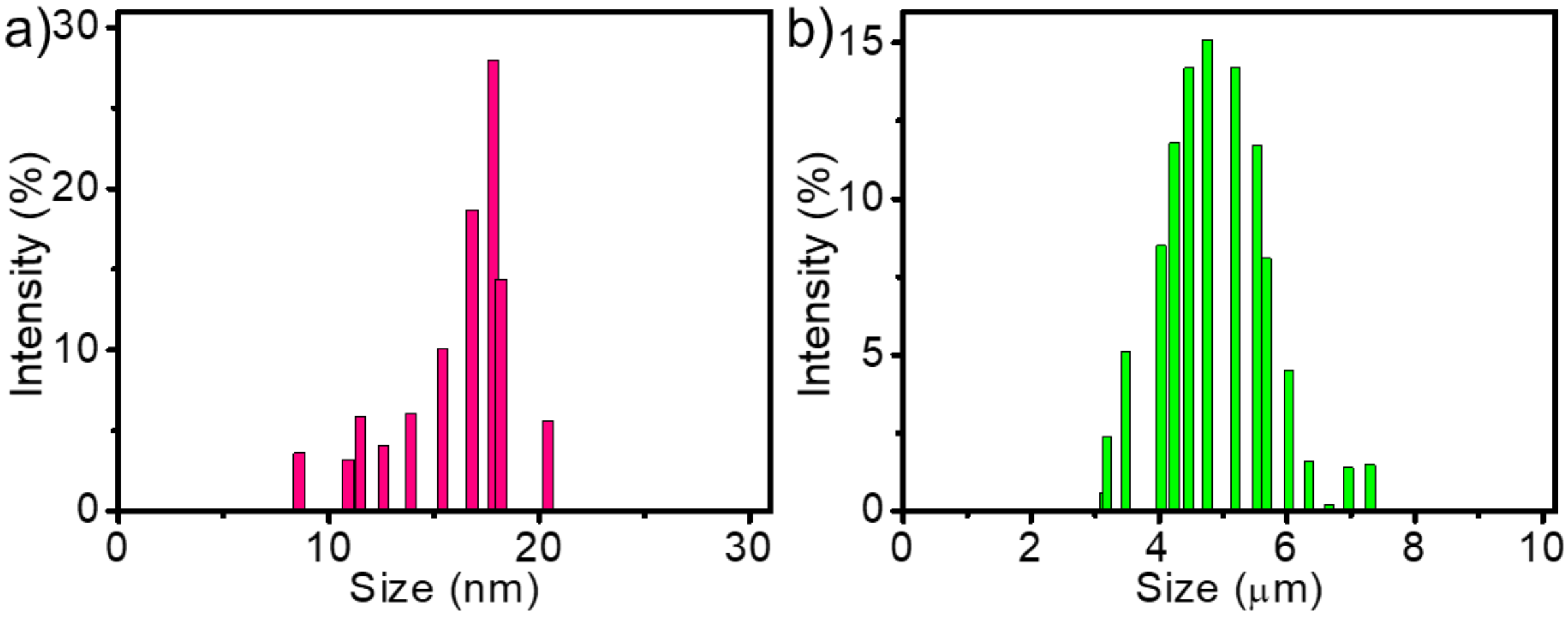

**Figure S4.** DLS spectra of colloidal $\delta$-$FAPbI_3$ a) NPs and b) NRs dispersed in toluene. The average size of the NPs is 17±2 nm, and the average length of the NRs is 5±1 m.

**6.** Conversion of NPs to NRs

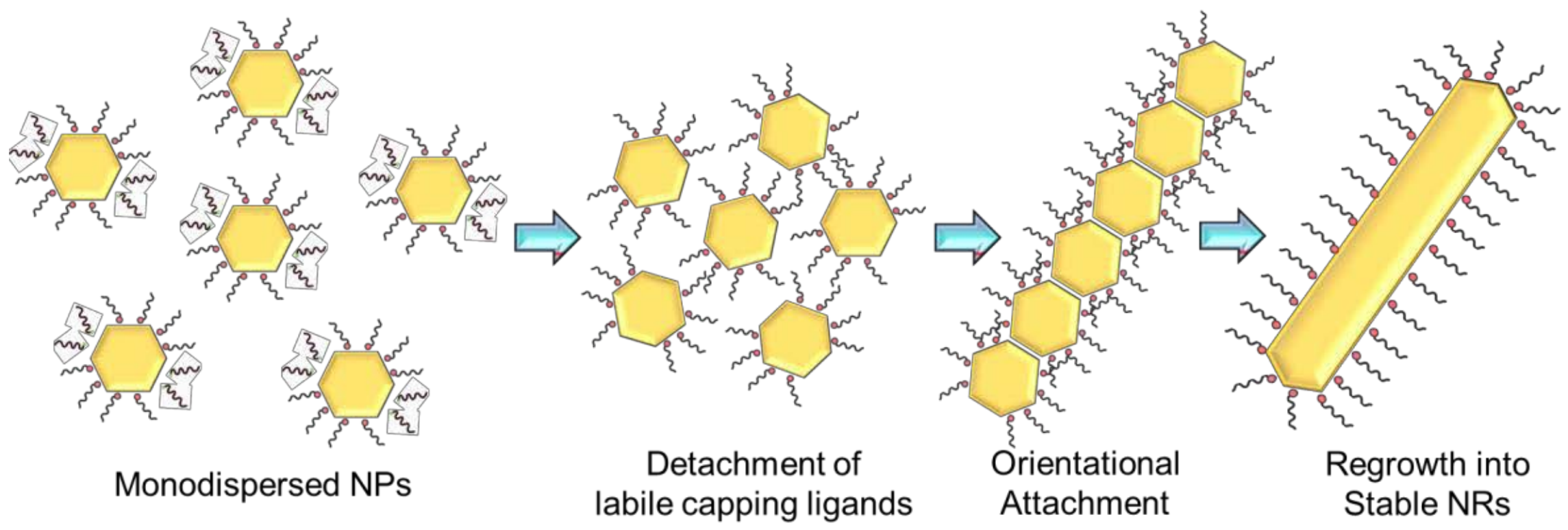

**Figure S5.** Schematic diagram representing the different processes involved in the formation of $\delta$-$FAPbI_3$ NRs.

**7.** Reflectance spectrum

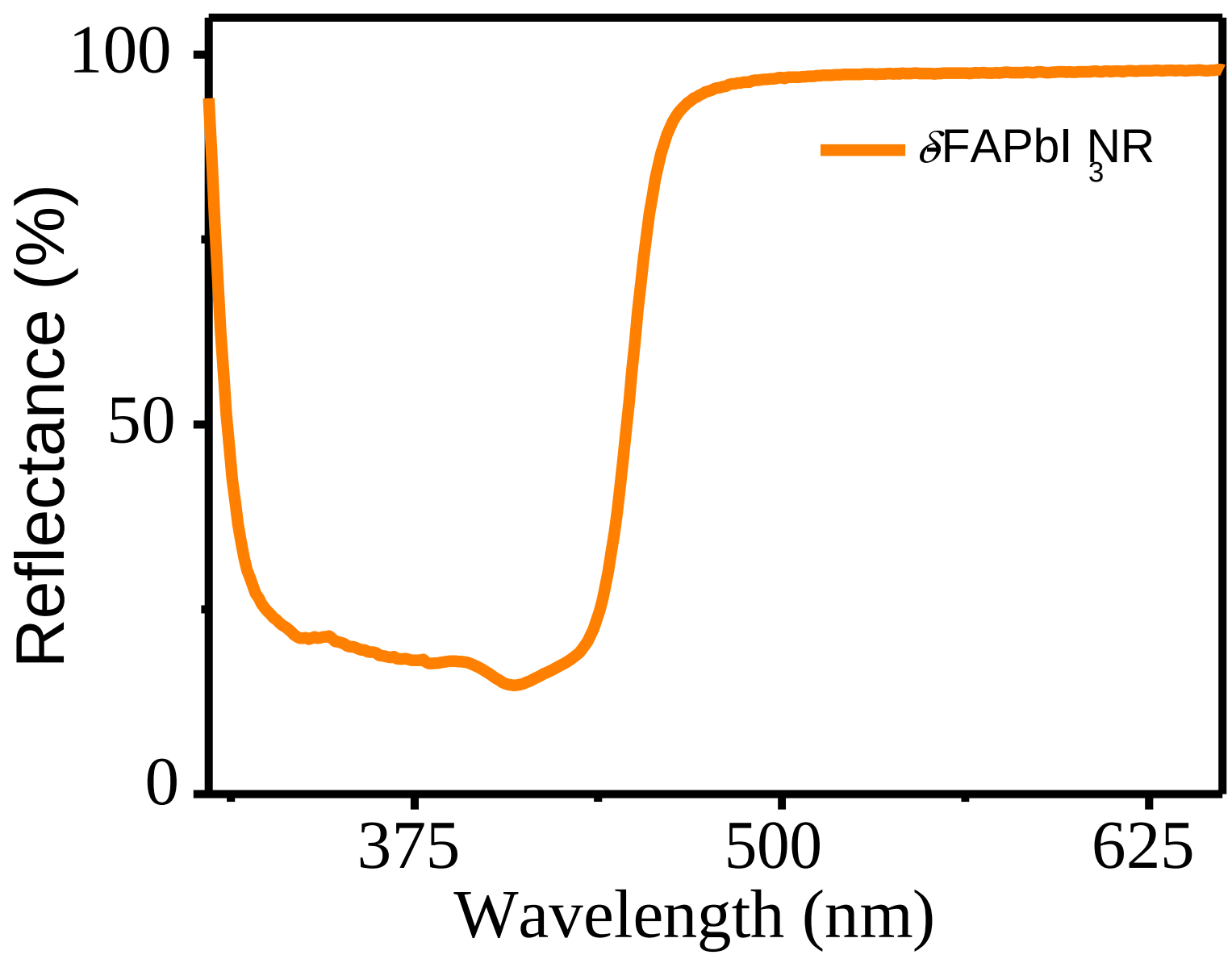

**Figure S6.** Reflectance spectrum of $\delta$-$FAPbI_3$ NRs film.

**8.** Weibull distribution

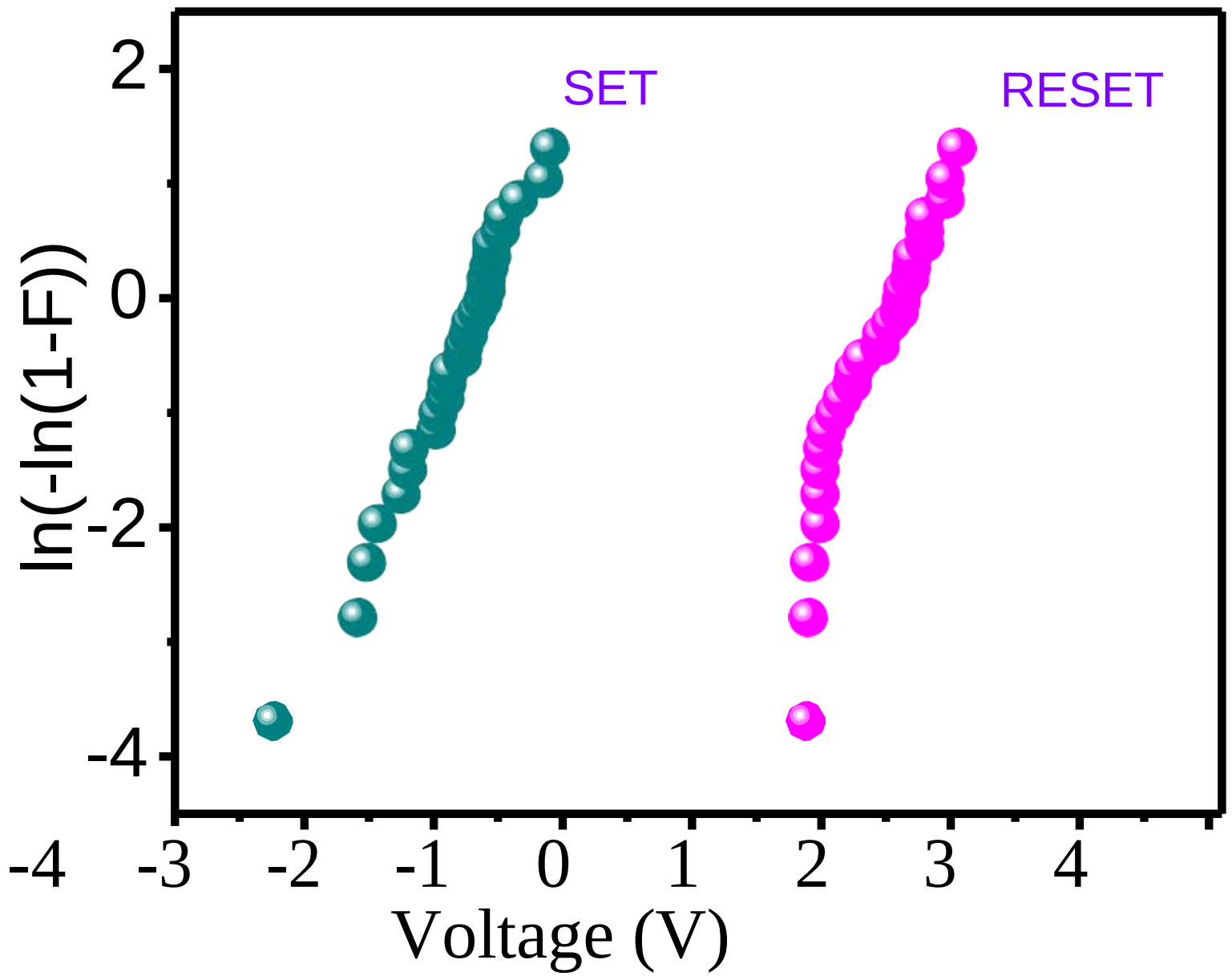

**Figure S7.** Weibull distributions of the SET and RESET voltages measured for 25 devices.

**9.** Endurance analysis

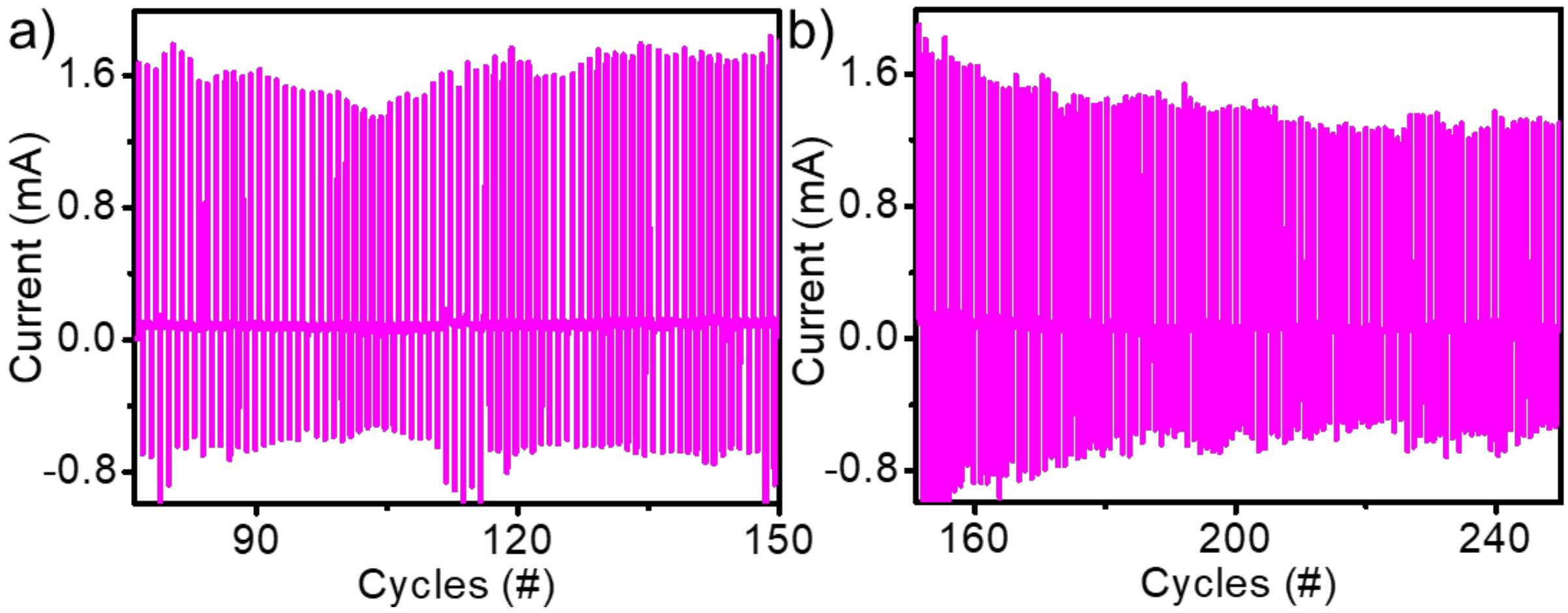

**Figure S8.** Cycling measurements of the device performed at pulse voltages of -3, 0.5, and 3V, respectively, for 250 cycles. The graphs are shown in two windows for clarity; however, the measurement data are continuous.

**10.** Stability studies

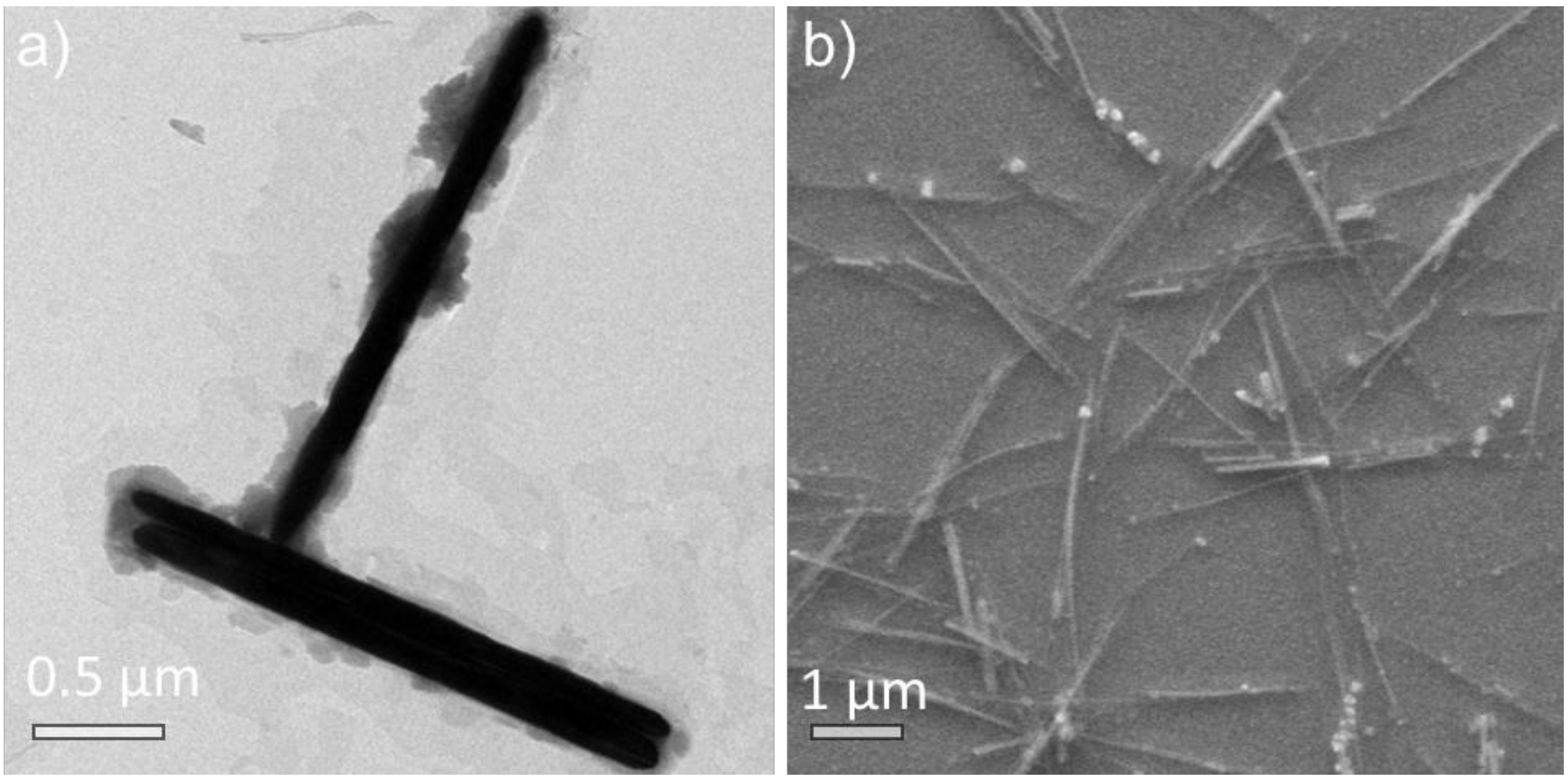

**Figure S9.** a) TEM and b) SEM images of $\delta$-FAPbI3 NRs taken after 12 months. The shape and size of the NRs remained largely unchanged, indicating long-term stability under ambient conditions.

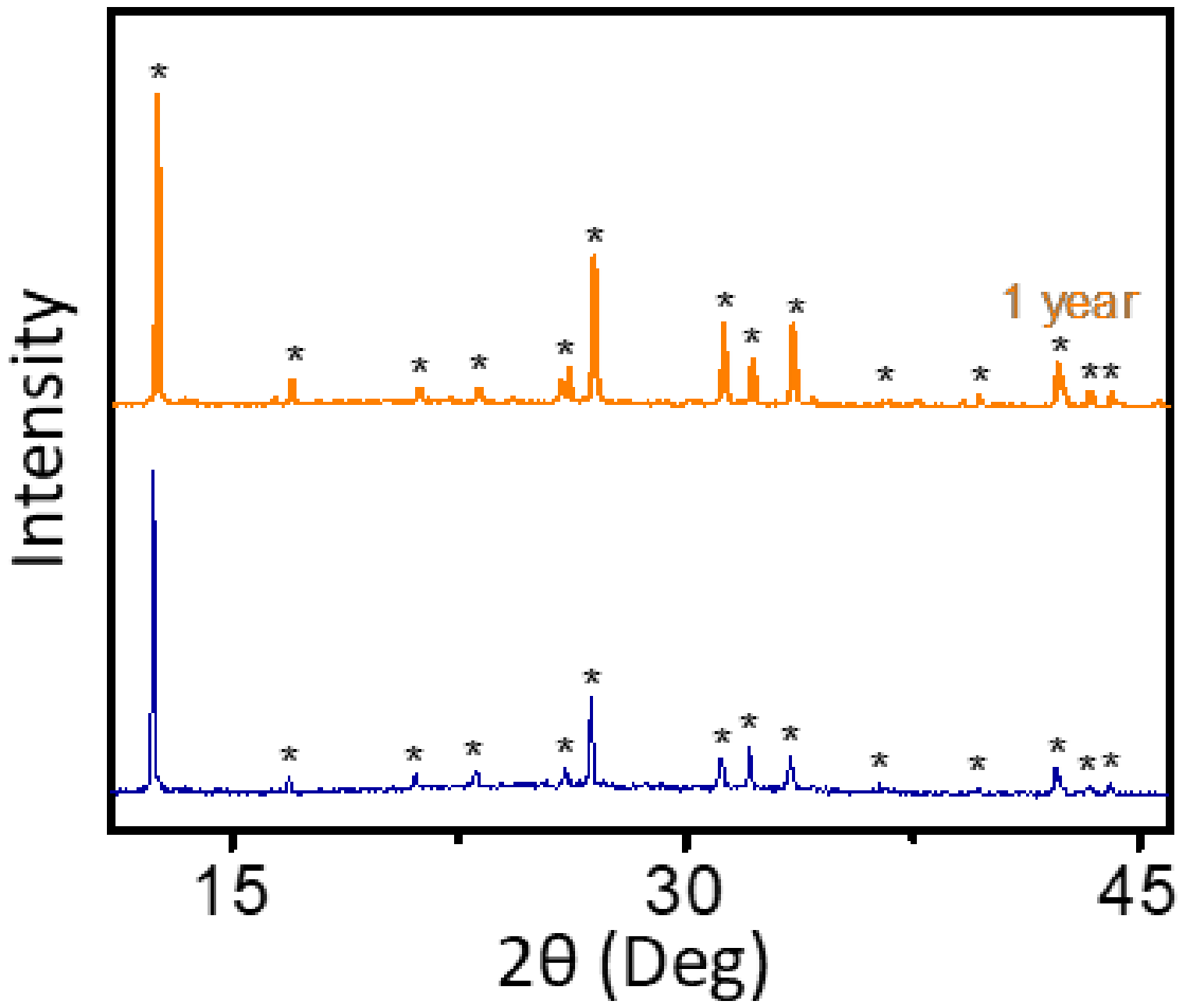

**Figure S10.** XRD spectrum of $\delta$-FAPbI3 NRs film measured on the first day (blue line) and after one year (orange line), confirming the long-term stability of the material.

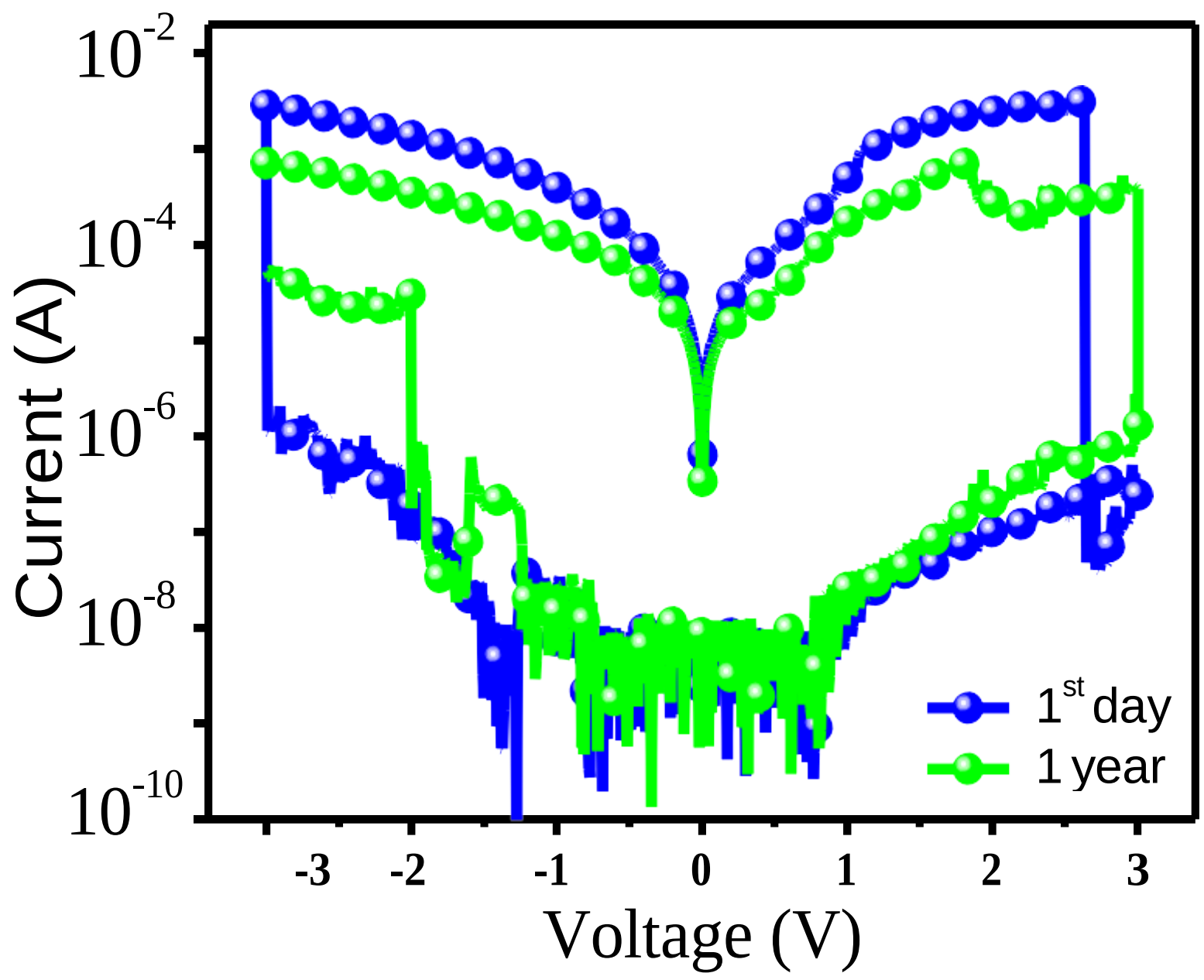

**Figure S11.** The bipolar resistive switching $\delta$-$FAPbI_3$ NRs film measured on the first day (blue line) and after one year (green line). The device exhibited similar resistive switching properties, confirming the long-term durability of the devices.

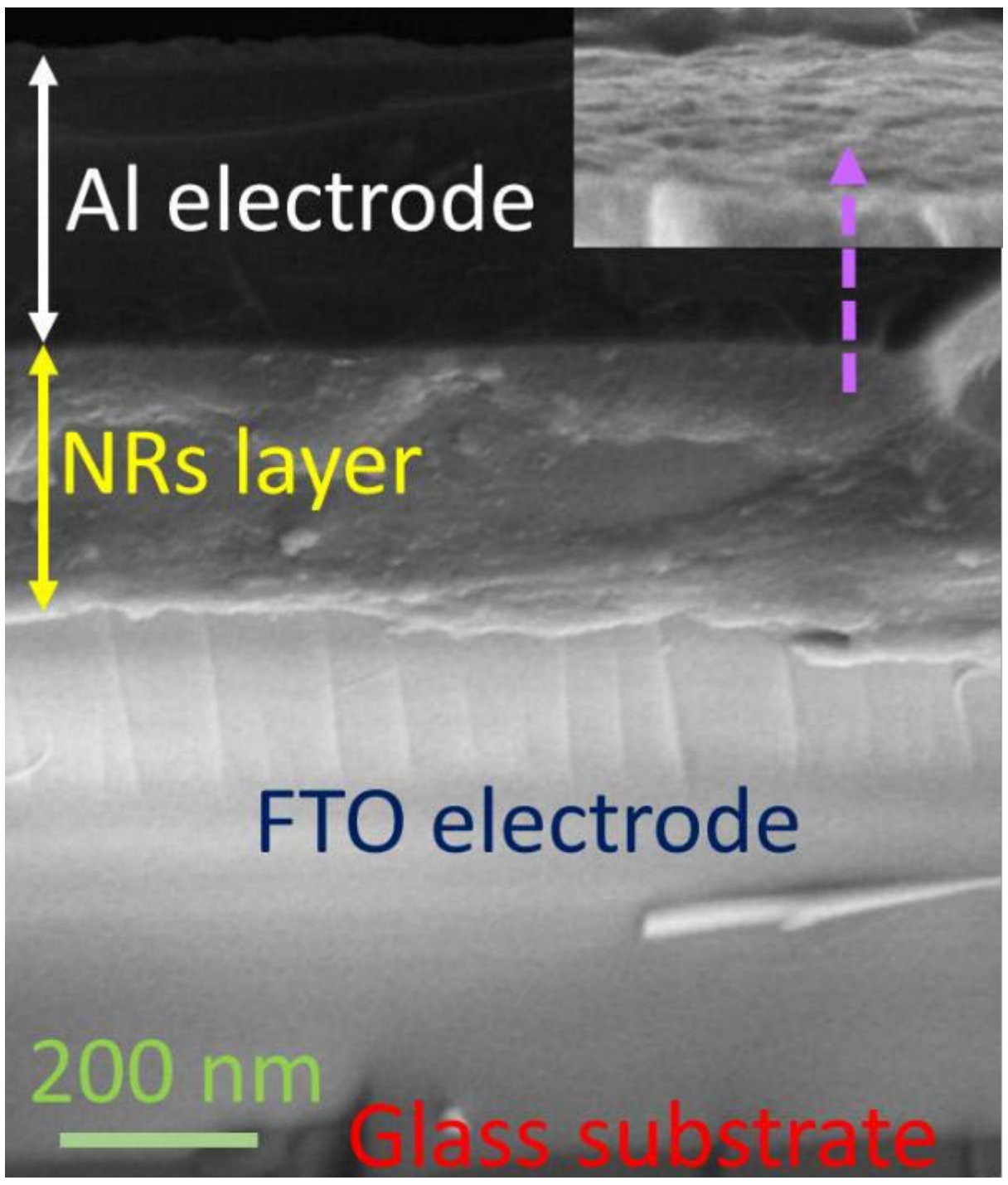

**Figure S12.** Cross-sectional SEM image of the FTO/$\delta$-$FAPbI_3$ NR/Al memory device after resistive switching and cycling studies. The active layer remained unaffected under the electric field, indicating the robustness of the material.

**11.** Device-level XPS in-depth analysis

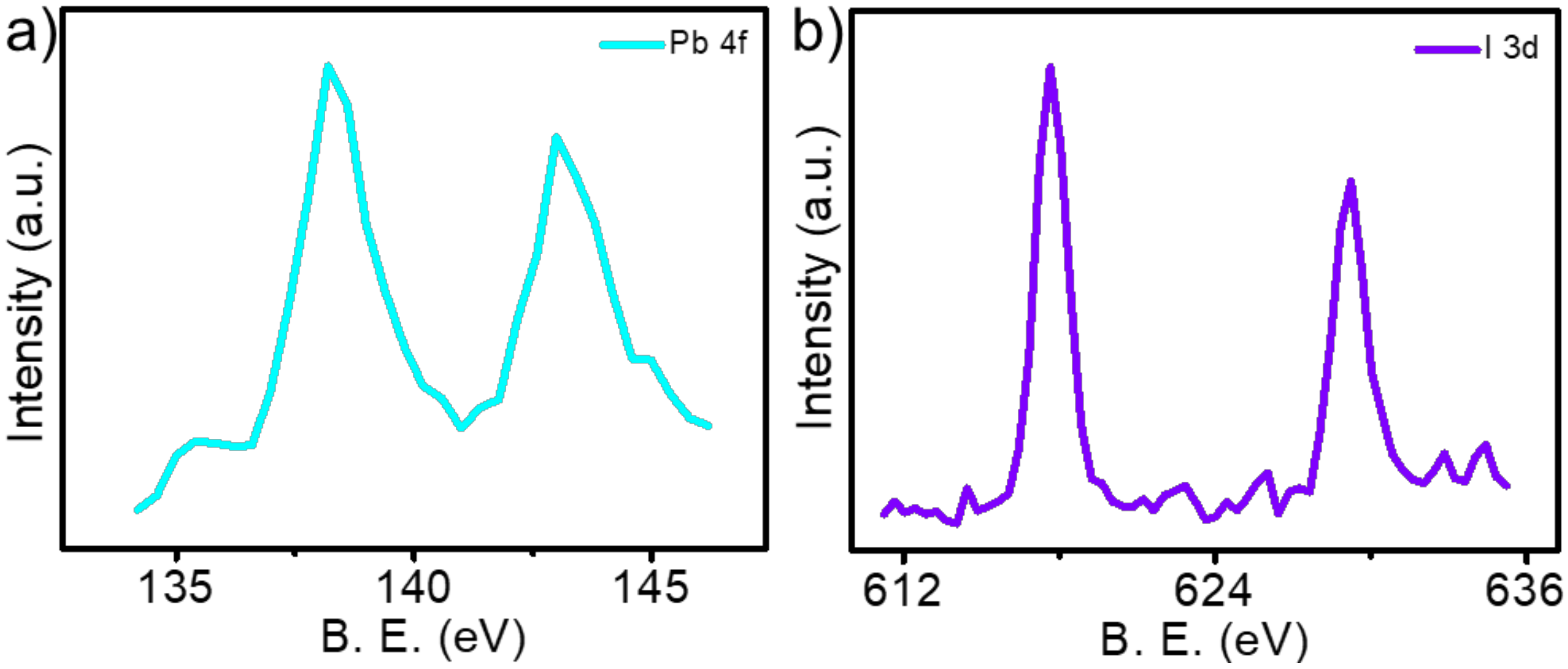

**Figure S13.** XPS profiles of NRs corresponding to a) Pb 4f and b) I 3d measured in the HRS (off-state).

**12.** XRD profile at LRS and HRS

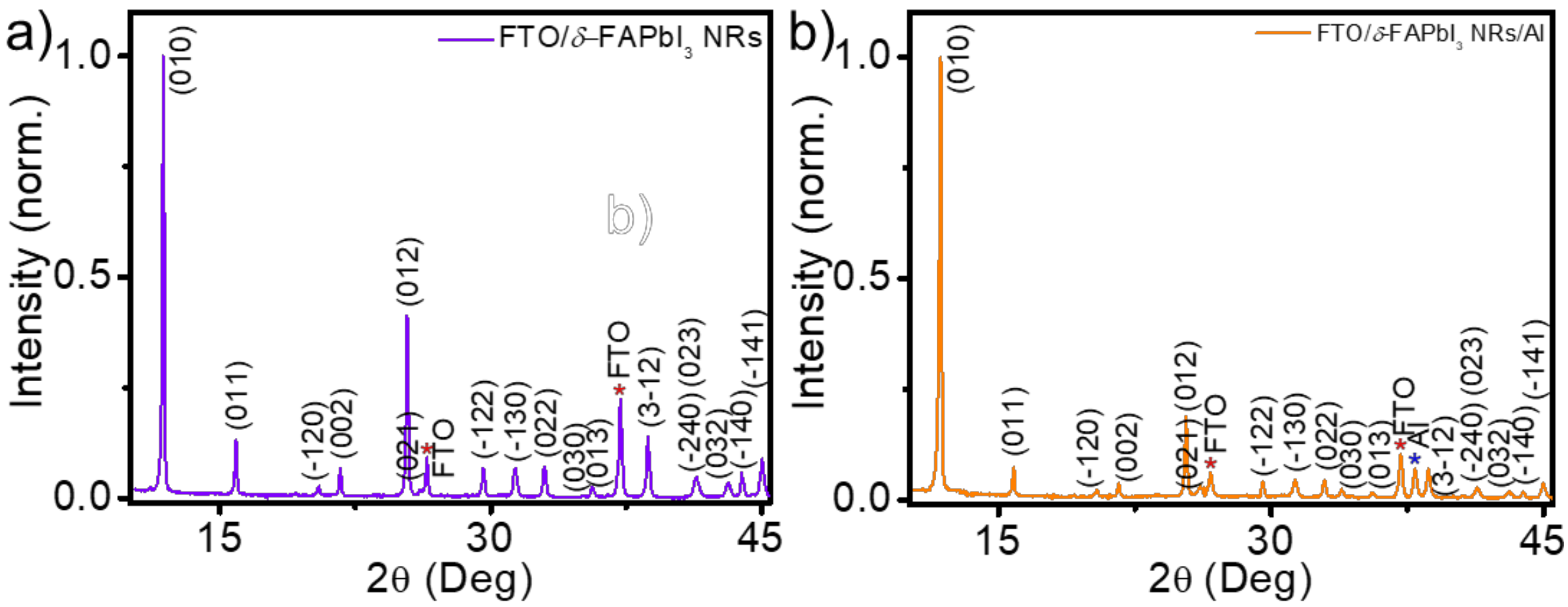

**Figure S14.** XRD spectra of $\delta$-FAPbI3 NRs active layer measured at a) LRS and b) HRS. The consistency in the peak positions and the absence of new peaks rule out phase transitions during resistive switching.

## 13. XPS analysis at different layer thickness

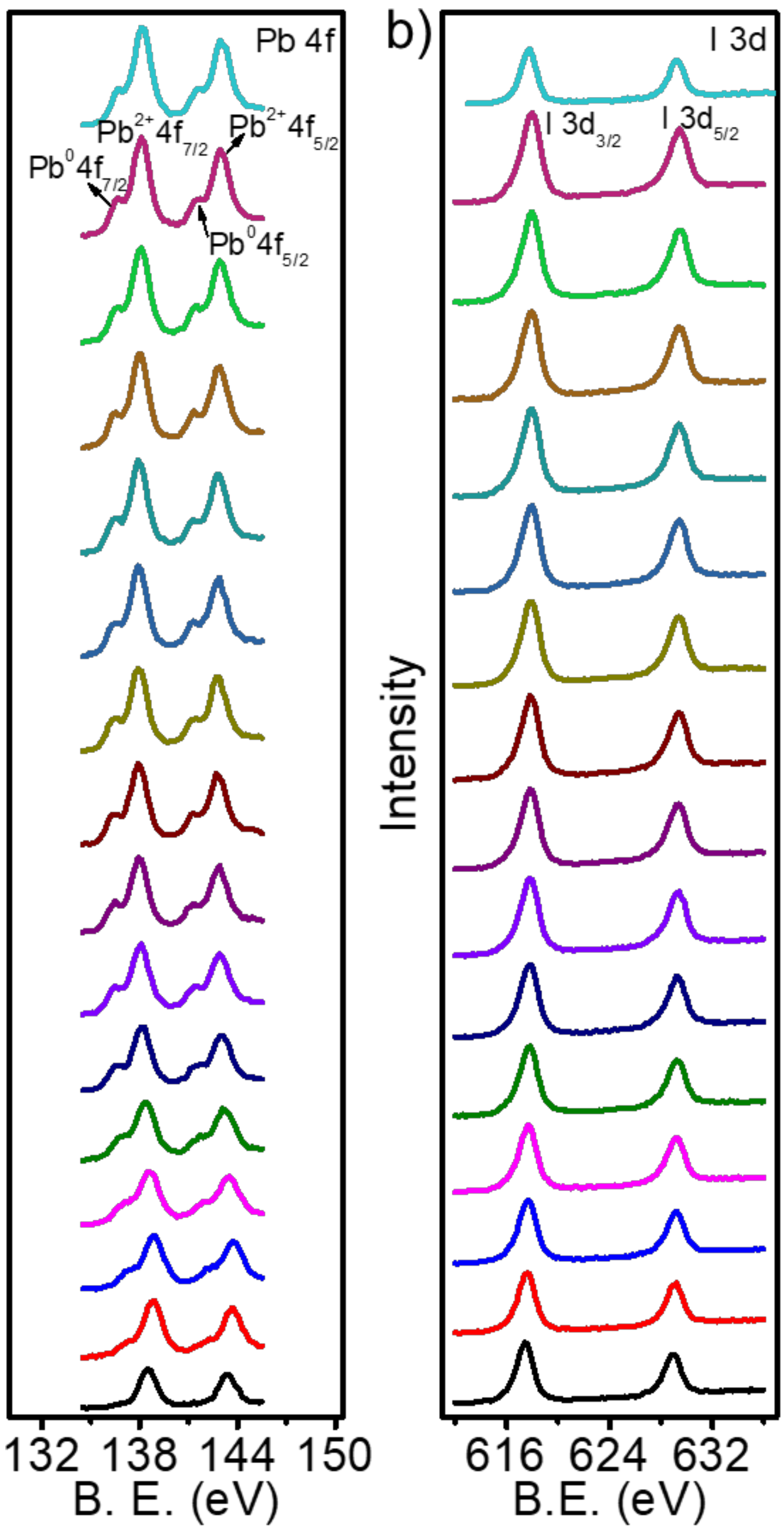

**Figure S15.** In-depth XPS profile of the $\delta$-FAPbI3 NRs film corresponding to a) Pb 4f and b) I 3d at different active layer thickness in the LRS. The presence of metallic lead in the different layers confirms the formation of halide ion vacancy filaments.

## 14. *I-V* measurement at orthogonal direction

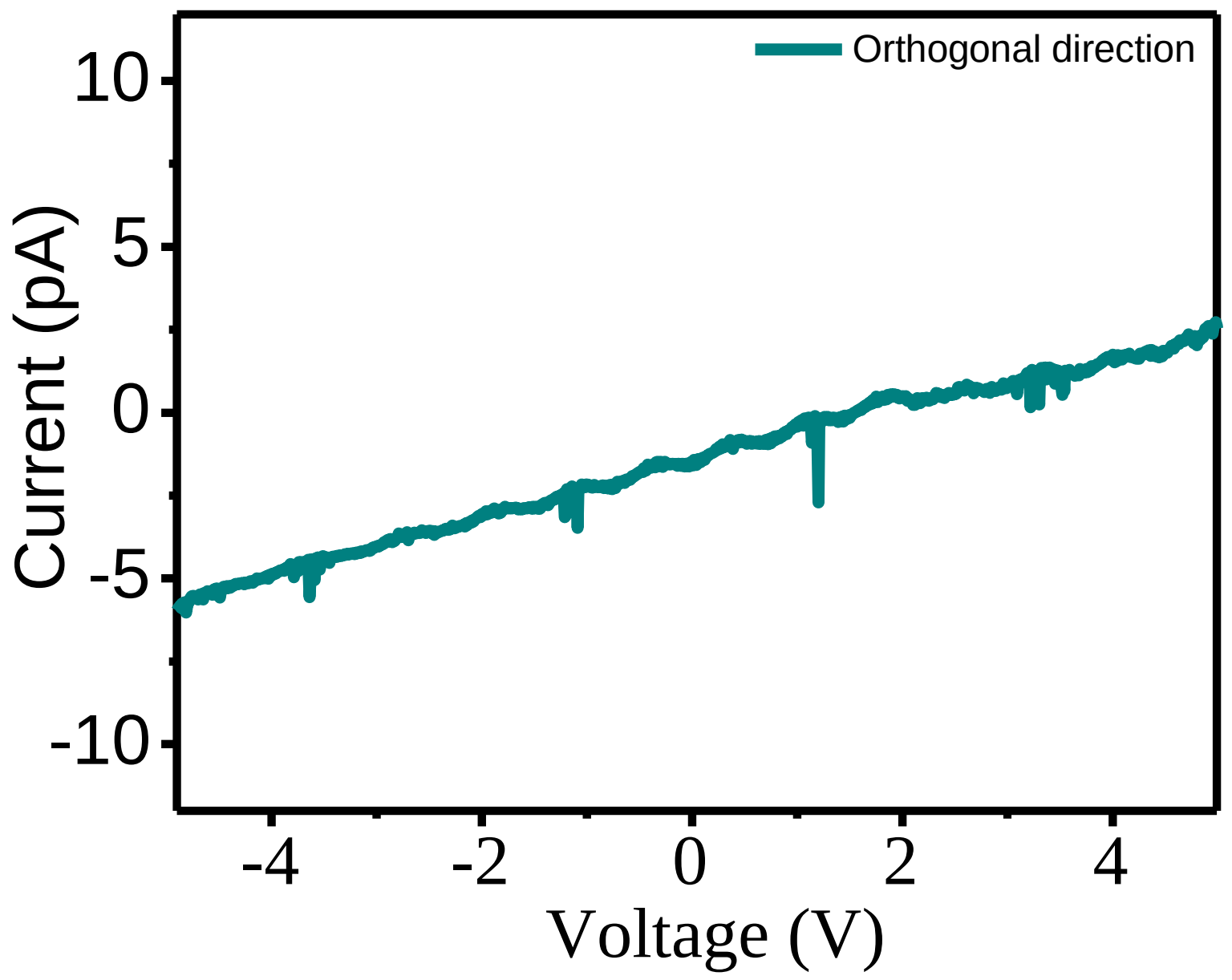

**Figure S16.** *I-V* curve for $\delta$-FAPbI3 NRs film, measured in the orthogonal direction to the electrodes, demonstrating poor conduction along the elongated axis of the NRs.

## 15. Schematic diagram showing orthogonal conduction

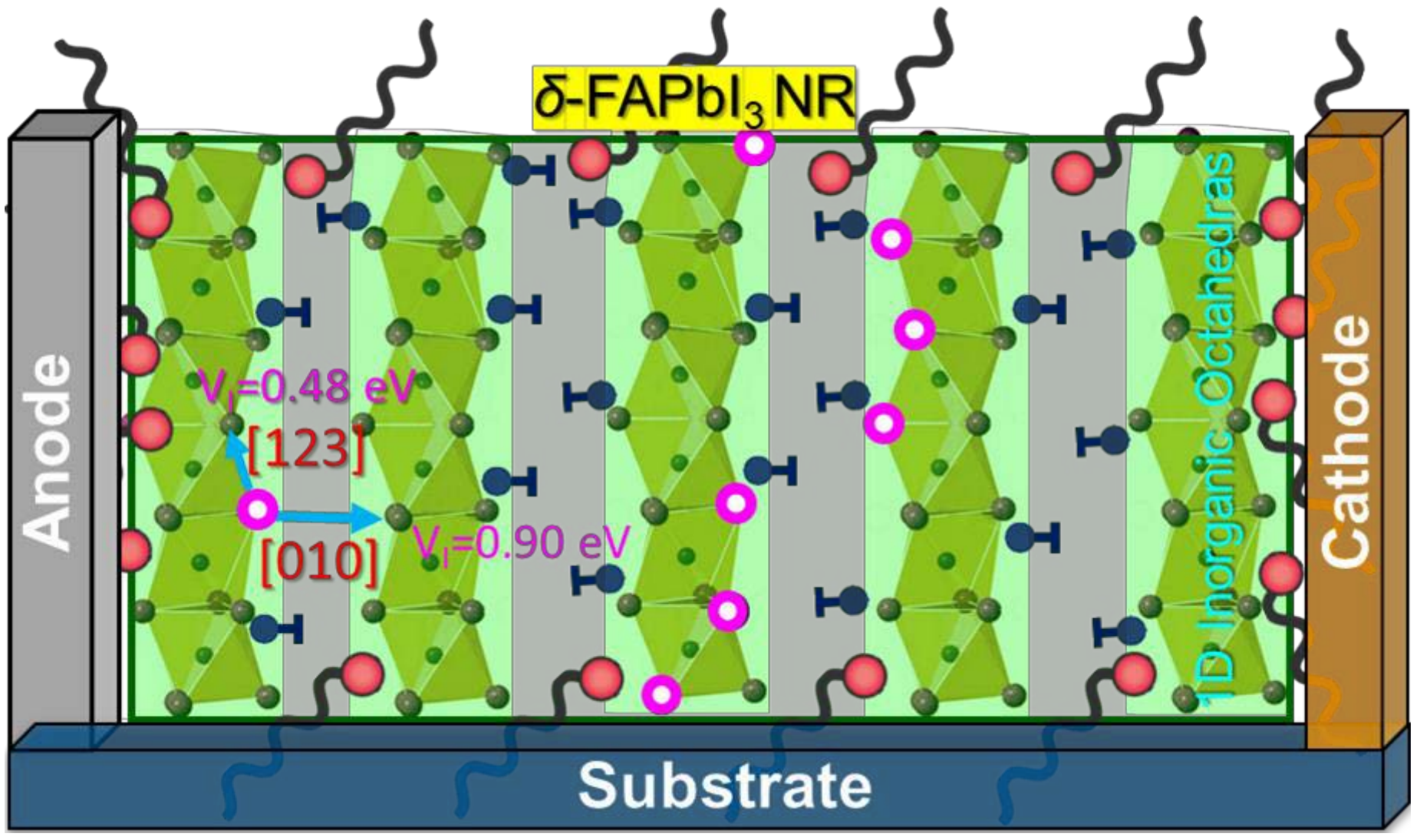

**Figure S17.** Schematic diagram illustrating a $\delta$-FAPbI3 NR orthogonally oriented towards the electrodes. The 1D inorganic $PbI_6^{4}$ octahedra within the NR were oriented parallel to the electrodes. The migration pathways of halide ion vacancies and their related migration energies are also shown.

## 14. *I-V* measurement at orthogonal direction

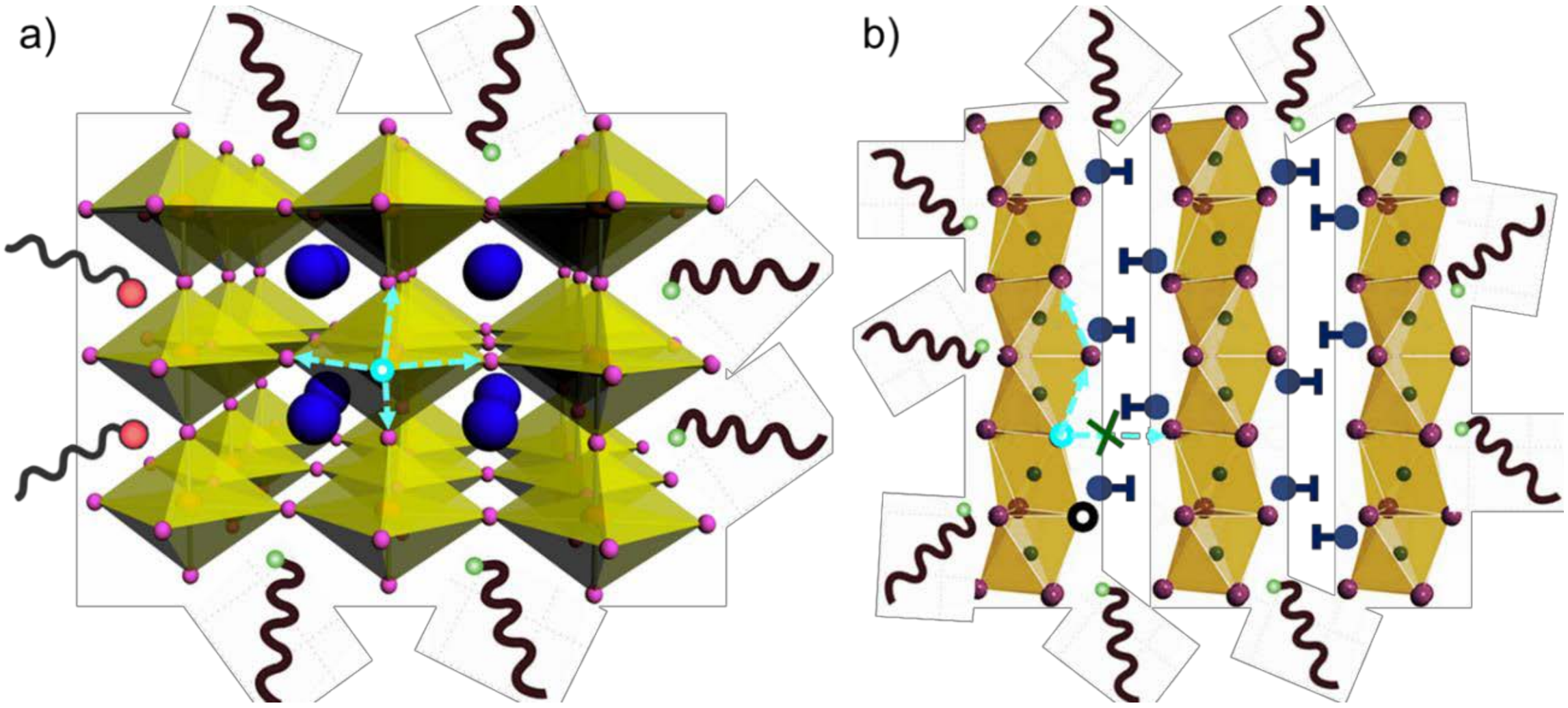

**Figure 18.** Schematic illustration of the different migration pathways of halide ion vacancies in a) $\alpha$-FAPbI3 NPs and b) $\delta$-FAPbI3 NRs. Halide ions can migrate in all directions within the 3D $\alpha$-FAPbI3 NP. The migration of halide ion vacancies is feasible along the 1D inorganic octahedra present in $\delta$-FAPbI3 NR.